\documentclass[9pt,twocolumn,twoside]{osajnl}

\journal{jocn} 

\setboolean{shortarticle}{false}
\title{Reinforcement Learning for Dynamic Resource Allocation in Optical Networks: Hype or Hope?}

\author[1*]{Michael Doherty}
\author[1]{Robin Matzner}
\author[1]{Rasoul Sadeghi}
\author[1]{Polina~Bayvel}
\author[1]{Alejandra~Beghelli}

\affil[1]{Optical Networks Group, University College London, Torrington Place, London WC1E 7JE, United Kingdom}

\affil[*]{Corresponding author: michael.doherty.21@ucl.ac.uk}

\begin{abstract}
    The application of reinforcement learning (RL) to dynamic resource allocation in optical networks has been the focus of intense research activity in recent years, with almost 100 peer-reviewed papers. We present a review of progress in the field, and identify weaknesses in benchmarking practices and reproducibility. To demonstrate best practice, we exactly recreate the problem settings from five landmark papers and apply improved benchmarks. To determine the best benchmarks, we evaluate several heuristic algorithms and optimize the candidate path count and sort criteria for path selection. We apply the improved benchmarks and demonstrate that simple heuristics outperform the published RL solutions, often with an order of magnitude lower blocking probability. Finally, to estimate the limits of improvement on the benchmarks, we present empirical lower bounds on blocking probability using a novel defragmentation-based method. Our method estimates that traffic load can be increased by 19--36\% for the same blocking in our examples, which may motivate further research on optimized resource allocation. We make our simulation framework and results openly available to promote reproducible research and standardized evaluation \hyperlink{https://doi.org/10.5281/zenodo.12594495}{https://doi.org/10.5281/zenodo.12594495}.%{10.5281/zenodo.12594495}.
\end{abstract}

\setboolean{displaycopyright}{false} % Do not include copyright or licensing information in submission.

\begin{document}

\maketitle

\section{Introduction}

% Is the motivation for applying RL to increase capacity or to increase uatoation? Both! But more so dynamicity and automation.

Network operators are faced with a problem: how to increase capacity for fast-growing data traffic without increasing the price of services \cite{lord_future_2021}. Advances in transmission technology have so far provided the solution by exponentially increasing the point-to-point capacity of the optical channel. However, as the throughput of the installed C+L bands approaches the nonlinearity-limited information bounds \cite{shtaif_information_2024}, costly infrastructure upgrades are required to scale capacity through spatial division multiplexing or ultra-wideband transmission \cite{winzer_future_2023}. Operators seek to minimize or delay the required capital expenditure and offset it with reduced operational costs. Online optimization of network resource allocation offers a path towards these aims by increasing the achievable network throughput with dynamic and automated service provisioning. 

Reinforcement learning (RL) has emerged as a promising technique for dynamic resource allocation (DRA) from a range of exact solution methods, heuristic algorithms, and artificial intelligence (AI) approaches.  RL solutions can approach the quality of exact methods such as integer linear programming (ILP), with an online allocation time comparable to simple heuristics \cite{di_cicco_deep_2022}. In fact, many works have demonstrated RL solutions that are superior to selected heuristics across various optical network problems %\cite{chen_deeprmsa_2019,chen_multi-task-learning-based_2021,xu_deep_2022} 
 (see section \ref{sec:survey}).

However, despite the many papers that investigate RL for optical networks, adoption of machine learning (ML) techniques by network operators has been limited by non-technological barriers \cite{khan_non-technological_2024}. One barrier is a lack of clear benchmarks and demonstrable benefits. Studies of resource allocation problems in optical networks often present results on specific topologies and traffic models, without fair comparison to previous results. Often results are not generalizable. 

In this paper we address this barrier by providing analysis of and recommendations for benchmarking practices. We identify 5 papers that provide the best examples of benchmarking in the field and exactly recreate their problem settings. By "problem settings", we mean the exact network topologies, traffic models, and other details of the network simulations. We then evaluate a range of heuristic algorithms and apply the best to each case. The best heuristic algorithms outperform or equal the reported RL results in all cases.

To understand if it is possible to improve on these new benchmarks, we propose a heuristic-based lower bound network blocking estimation method, termed Resource-Prioritized Defragmentation (section \ref{sec:bounds}). For each case of study, we apply this method over a range of traffic loads and estimate that the supported traffic load at a fixed blocking (0.1\%) may be increased by 19-36\% for flex-grid networks. These estimates suggest further improvement is possible.

This is the first time that a thorough analysis and benchmarking of previous work has been carried out, and highlights deficient benchmarking standards. Our findings suggest that previous attempts to find better resource allocation policies using RL have been unsuccessful, and new efforts are required to find solutions, using RL or other methods, that improve on the best benchmarks and approach the estimated bounds.

%We also investigate the limits of network throughput for dynamic traffic, to understand how much additional data traffic could be supported by improved resource allocation. We propose a heuristic-based lower bound network blocking estimation method, termed Resource-Prioritized Defragmentation (section \ref{sec:bounds}). For each case of study, we apply this method over a range of traffic loads and estimate that the supported traffic load at a fixed blocking (0.1\%) may be increased by 19-36\% for flex-grid networks. These estimates may motivate further research on optimized DRA with RL or other methods.

This paper aims to: 1) Promote higher standards for evaluation and benchmarking in research on RL for DRA. 2) Improve reproducibility and transparency of research by open sourcing our simulation framework and providing discussion of implementation details that significantly affect results (e.g. path ordering in Section 4.\ref{sec:path_ordering} and holding time truncation in Section 5.\ref{sec:holding_time}). 3) Encourage new research into optimized DRA by demonstrating that previous RL approaches have failed to beat our benchmarks but there remains a considerable optimality gap between the benchmarks and our estimated bounds.

The contributions of this work are:
\begin{enumerate}[itemsep=0pt]
    \item A comprehensive survey of progress on RL approaches to DRA in optical networks.
    \item A systematic study of the factors affecting heuristic algorithm performance and recommendations for better heuristic benchmarks.
    \item A recreation of problems from five landmark papers in the field with comparisons to improved heuristic benchmarks, showing previous RL results have failed to improve on heuristics.
    \item The introduction of a novel empirical throughput bound estimation method, which shows benchmarks can be improved.
    \item The release of our simulation framework, "XLRON", to promote reproducible research and enable fair comparison across different studies.
\end{enumerate}

All of the code necessary to generate data and plots from this paper are available on Github \cite{michael_doherty_micdohxlron_2024}.

We begin with essential background on DRA problems and RL techniques in Section 2, followed by a literature review in Section 3. Section 4 presents the investigation of heuristic algorithms for benchmarking. Section 5 details the analysis of the previously published results and comparison with benchmarks. Section 6 presents new empirical bounds for network throughput, with recommendations for future research directions in Section 7.

\section{Background}
\label{sec:background}

\subsection{DRA problems in optical networks}

%\subsubsection*{Network Capacity and Low-Margin Operation}
%The optical data communication infrastructure that underpins metro, inter-data center, national and continental scale networks presents a range of optimization problems. The objective of these problems is to maximize the network throughput (the sum of realized point-to-point data transmission connections) within the constraints of available resources, most crucially the available bandwidth on links.

%The complexity of these systems has grown as networks have become increasingly flexible in the parameters of each transmitted channel; including launch power, symbol rate, modulation format and channel spacing. This flexibility is enabled by now-ubiquitous coherent transponders with advanced DSP, and ROADMs that support varying spectrum granularity. The highly configurable nature of each channel opens opportunities to increase network utilization through low-margin operation \cite{auge_can_2013}. Unallocated margins and system margins \cite{pointurier_design_2017} can be reduced with this technology by matching the requested and available channel capacity and reach, and mitigating non-linear effects through launch power allocation, respectively. Moving from fixed-grid (50GHz or greater channel spacing) to flex-grid (minimum 6.25GHz channel) spacing allows more precise matching of requested data transmission capacity to bandwidth. Network capacity is further enhanced by judicious allocation of routes and bandwidth to network traffic, which is the primary task of DRA problems.

\subsubsection*{Motivation for dynamic operation and RL}
As discussed by Aug\'e \cite{auge_can_2013} and Pointurier \cite{pointurier_design_2017}, optical networks must operate with a margin of additional resources between the minimum requirements to fulfill a data service request and the resources allocated to that request, for example by allocating additional spectrum. This margin is required due to uncertainty in physical parameters of the transmission network or future traffic variations. Reducing the margin can increase the network throughput or reduce costs.

Dynamic operation allows the allocated resources to vary temporally in response to shifting traffic, thereby allowing more accurate matching of resources to current demand, which reduces margins. The time constraints of dynamic operation may preclude exact solution methods, but a trained RL agent can compute an allocation in sub-second time \cite{di_cicco_deep_2022}.

RL is appropriate for these problems because they possess three characteristics that merit the application of AI \cite{hassabis_nobel_2024}: 1) a large combinatorial search space, 2) a clear objective function for optimization, and 3) plentiful training data and/or accurate and efficient simulators for data generation. 
%\cite{nevin_techniques_2022,di_cicco_deep_2022}.

\subsubsection*{Traffic models}
Network traffic comprises a set of requests to connect source and destination nodes with fixed data rates on dedicated lightpaths. Traffic can be modeled as static, incremental, or dynamic \cite{zang_review_2000}.

Static traffic assumes knowledge of all connection requests that the network must accommodate. Incremental traffic lies between static and dynamic in stochasticity: requests are not known in advance but do not expire once allocated. For dynamic traffic, connection requests are served on-demand without knowledge of future requests, and active connections expire randomly. The request arrival and expiry times are sampled from probability distributions, often assumed to be exponential \cite{chen_deeprmsa_2019}. Dynamic traffic is considered a paradigm for future optical networks, that have the necessary systems in place to enable real-time response to changing network conditions \cite{lord_flexible_2022}.

%In current production networks, traffic demand is typically forecast for a given period e.g. one year, and resources allocated and configured in advance. Additional unforecast demand may then be added incrementally. Dynamic traffic is therefore a paradigm for future optical networks, that have all the necessary software and hardware systems in place to enable real-time response to changing network conditions  \cite{lord_flexible_2022}.
%In this paper, we consider only dynamic traffic, as this paradigm offers the , though we note that previous research into wavelength-routed networks found dynamic operation only increased total capacity in networks with wavelength conversion capability \cite{zapata-beghelli_dynamic_2008}.

%The traffic model spatial characteristics are defined by the traffic matrix. %In dynamic or incremental settings, the traffic matrix elements are normalized probabilities of a connection between the nodes corresponding to the column and row indices \cite{nevin_techniques_2022}. %In static settings, a traffic matrix element may represent the total data rate between the node pair \cite{jaumard_decomposition_2023}. %It is common practice to consider uniform traffic matrices for DRA problems.

\subsubsection*{Problem variants}
The classic optimization problem in an optical network is Routing and Wavelength Assignment (RWA) for fixed-grid networks or Routing and Spectrum Assignment (RSA) for flex-grid networks, where spectrum is divided into frequency slot units (FSU) \cite{mukherjee_springer_2020}. Further degrees of freedom in the optimization are added by considering the selection of \underline{M}odulation formats (R\underline{M}SA), and the fiber \underline{C}ore or spectral transmission \underline{B}and utilized by each channel in the case of multi-core or multi-band networks (R\underline{C}MSA/R\underline{B}MSA). Launch \underline{P}ower has also been considered as a parameter in the optimization objective for dynamic networks (R\underline{P}MSA) \cite{ives_routing_2015} \cite{arpanaei_launch_2023}.% and is a relatively under-investigated problem variant.  

While most DRA problems in optical networks from the literature are concerned with point-to-point connections, some consider virtual networking tasks such as virtual optical network embedding (VONE) \cite{gong_virtual_2014,doherty_deep_2023} or virtual network function placement (VNF) \cite{zhou_applications_2022}. We choose to focus on RWA/RSA/RMSA in this paper because they are the most widely studied DRA problems in the context of optical networks and they form a core sub-task of variants such as VNF or VONE.

% Overview paper of problem types (also inlcudes details on ML/RL but very limited and with poor understanding): \cite{zhang_overview_2020}

\subsubsection*{Constraints}
Three fundamental constraints govern resource allocation in optical networks:
%transparent\footnotemark optical networks:

%\footnotetext{This assumes lightpaths are routed "transparently" at nodes without optoelectronic conversion for the purposes of signal regeneration or wavelength conversion.}

\begin{enumerate}[itemsep=0pt]
    \item Spectrum Continuity: A lightpath must use identical FSU on each link, without wavelength conversion.
    \item Spectrum Contiguity: FSU allocated to a lightpath must be adjacent.
    \item No Reconfiguration: Active lightpaths cannot be reallocated once established, meaning allocation decisions are permanent while connections remain active.
\end{enumerate}

The 'No Reconfiguration' constraint is not a physical limitation but an operational assumption that active services cannot be disrupted. Reconfiguration may sometimes be desirable, especially in flex-grid networks that may suffer from spectral fragmentation \cite{gerstel_elastic_2012}, and has been used in a production network by Meta Platforms Inc. to free up spectral resources \cite{balasubramanian_targeted_2023}.

\subsubsection*{Solution methods}
%The possible methods to solve the DRA problems depend on whether the traffic is static or dynamic. Static traffic allows analytical approaches for exact combinatorial optimization, most notably Integer Linear Programming (ILP). ILP formulations have been suggested for DRA problems \cite{walkowiak_ilp_2016}, but the application of ILP to realistic problem sizes of networks with capacity for thousands of connection requests remains computationally infeasible \cite{jaumard_decomposition_2023}.% The RWA problem for static traffic has been shown to be NP-hard \cite{chlamtac_lightpath_1992}. Consequently, the computational complexity scales super-polynomially with the topology size and traffic load for RWA and related problems.

Allocation of static traffic is a NP-hard combinatorial optimization problem \cite{chlamtac_lightpath_1992}, for which the computational complexity of finding a solution scales super-polynomially with the space of possible allocations. Exact solution methods such as Integer Linear Programming (ILP) have been formulated for static traffic \cite{walkowiak_ilp_2016,jaumard_decomposition_2023} but show limited scalability\footnotemark and are infeasible for dynamic traffic without a priori knowledge of all requests. \footnotetext{\cite{jaumard_decomposition_2023} Jaumard et al. scale their ILP formulation to 690 requests on the USNET topology (24 nodes and 86 links) with 380 FSU per link, without considering distance-adaptive modulation formats.}

For dynamic traffic, any allocation decision must be taken within the constraint of the time interval between request arrivals. No standard limit has been defined for this constraint in the literature but it could be on the order of seconds, with a lower limit set by the switching time of reconfigurable optical add-drop multiplexers (currently 1ms to 100ms, depending on the switching technology \cite{goto_lcos-based_2024}).

Many simple heuristic algorithms \cite{vincent_scalable_2019,abkenar_best_2016,wright_minimum-_2015,tang_heuristic_2022,savory_congestion_2014} have been proposed for these problems, with the goal to minimize resources, lightpath distances and required bandwidth. Heuristics have the advantages of fast execution time and deterministic and interpretable allocation decisions.

Machine learning approaches to DRA problems have included nature-inspired techniques like particle swarm optimization (PSO) \cite{hassan_chaotic_2009} and genetic algorithms (GA) \cite{barpanda_genetic_2011}. Once trained, an RL policy can compute an allocation faster than other ML approaches \cite{di_cicco_deep_2022}.

\subsection{Reinforcement learning}
\label{sec:background_rl}

RL is a framework for learning to optimize sequential decision making under uncertainty. It emerged from Bellman's foundational work on optimal control theory in the 1950s, specifically dynamic programming and Markov Decision Processes (MDP) \cite{bellman_theory_1954,bellman_markovian_1957}. MDPs model decision-making processes as an agent that takes actions in an environment to maximize a reward signal. The method of action selection is a mapping from states to actions termed the policy, which can be expressed in tabular form or approximated by a neural network (NN). The use of NN for function approximation is sometimes distinguished as "Deep" RL. Tabular RL has been applied to optical networks \cite{terki_routing_2023}, but function approximation with NN is widely used \cite{chen_deeprmsa_2019,shimoda_mask_2021,tang_heuristic_2022,xu_deep_2022,cheng_ptrnet-rsa_2024} when the set of state-action pairs is too large to be tabulated \cite{sutton_reinforcement_2018}.

The reader is referred to Sutton and Barto's authoritative textbook \cite{sutton_reinforcement_2018} for details. However, to aid the discussion of RL applied to DRA problems, several terms are defined here.

RL algorithms can be classified as model-based or model-free. Model-based RL uses a model of the environment to plan future actions, but so far no works have applied this paradigm to DRA problems in optical networks. Model-free RL algorithms learn through direct interaction with the environment, without planning. Model-free algorithms can be further categorized into:
\begin{itemize}
\item \textbf{Action-value methods}, such as Q-learning, which learn to estimate the value of taking actions in different environment states. These methods were the first to be developed in RL \cite{watkins_learning_1989,sutton_learning_1988} and have been used for route selection in optical networks \cite{bryant_q-learning_2022}.
\item \textbf{Policy gradient methods}, which directly optimize the policy parameters to maximize expected rewards. These methods can handle continuous action spaces, unlike action-value methods, which optimize an action-value function. Policy gradient methods are enhanced by using an Actor-Critic architecture, as in algorithms such as A2C \cite{mnih_asynchronous_2016} and PPO \cite{schulman_proximal_2017}, which reduce variance in the policy gradient by using a learned value function (critic) to estimate the value of each state. Policy gradient methods have been used in many works on DRA in optical networks, such as A2C for DeepRMSA \cite{chen_deeprmsa_2019}. 
\end{itemize}

\section{Literature Survey}
\label{sec:survey}

There exists a considerable body of literature on RL for DRA problems in optical networks. %\cite{amin_survey_2021}. 
DRA in optical networks is distinguished from similar problems in electronically linked networks by the nature of fiber optic links, which carry a set of wavelengths or FSU, defined by the ITU standards G.671 and G.694.2 \cite{international_telecommunication_union_spectral_2002,international_telecommunication_union_transmission_2012}. In this work we only consider publications related to optical networks but acknowledge the closely related literature on RL for other graph-based resource allocation problems.

\begin{figure}
    \centering
    \includegraphics[width=1\linewidth]{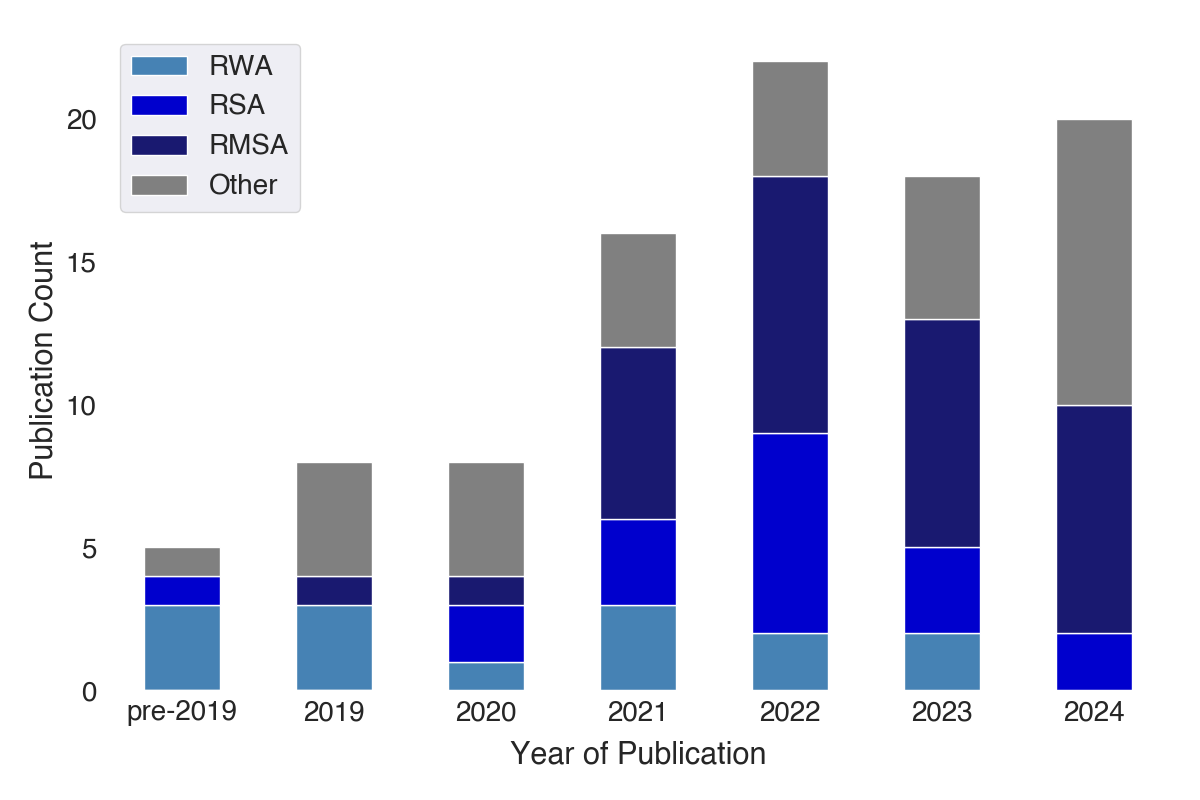}
    \caption{Count of publications related to RL for resource allocation problems in optical networks. Citations for each classification category are: RWA \cite{garcia_multicast_2003,pointurier_reinforcement_2007,koyanagi_reinforcement_2009,suarez-varela_routing_2019,shiraki_dynamic_2019,shiraki_reinforcement-learning-based_2019,tzanakaki_self-learning_2020,zhao_cost-efficient_2021,freire-hermelo_dynamic_2021,liu_waveband_2021,nevin_techniques_2022,di_cicco_deep_2022,di_cicco_deepls_2023,nallaperuma_interpreting_2023}, RSA \cite{reyes_adaptive_2017,li_deepcoop_2020,li_multi-objective_2020,romero_reyes_towards_2021,zhao_reinforced_2021,wang_dynamic_2021,quang_magc-rsa_2022,cruzado_reinforcement-learning-based_2022,zhao_rsa_2022,jiao_reliability-oriented_2022,almasan_deep_2022,wu_service_2022,arce_reinforcement_2022,sharma_deep_2023,lin_deep-reinforcement-learning-based_2023,hernandez-chulde_experimental_2023,cheng_ptrnet-rsa_2024,chen_gsaddqn_2024}, RMSA \cite{chen_deeprmsa_2019,wang_resource_2020,shimoda_mask_2021,shi_deep-reinforced_2021,shimoda_deep_2021,sheikh_multi-band_2021,xu_spectrum_2021,chen_multi-task-learning-based_2021,gonzalez_improving_2022,bryant_q-learning_2022,terki_routing_2022,tang_deep_2022,cheng_routing_2022,xu_deep_2022,tang_heuristic_2022,tu_entropy-based_2022,momo_ziazet_deep_2022,pinto-rios_resource_2023,errea_deep_2023,beghelli_approaches_2023,terki_routing_2023,tanaka_pre-_2023,xu_hierarchical_2023,sadeghi_performance_2023,tang_routing_2023,teng_deep-reinforcement-learning-based_2024,xiong_graph_2024,teng_drl-assisted_2024,unzain_reinforcement_2024,zhou_opti-deeproute_2024,li_opticgai_2024,xie_physical_2024,yan_drl-based_2024}, Other \cite{boyan_packet_1993,ma_demonstration_2019,zhao_reinforcement-learning-based_2019,wang_subcarrier-slot_2019,luo_leveraging_2019,natalino_optical_2020,ma_co-allocation_2020,wang_deepcms_2020,weixer_reinforcement_2020,liu_multi-agent_2021,zhao_service_2021,tian_reconfiguring_2021,morales_multi-band_2021,tanaka_reinforcement-learning-based_2022,koch_reinforcement_2022,hernandez-chulde_evaluation_2022,etezadi_deepdefrag_2022,etezadi_deep_2023,tanaka_adaptive_2023,johari_drl-assisted_2023,zhang_admire_2023,fan_blocking-driven_2023,lian_dynamic_2024,li_tabdeep_2024,wang_availability-aware_2024,yin_dnn_2024,tse_reinforcement_2024,tanaka_reinforcement-learning-based_2024,doherty_xlron_2024,natalino_optical_2024,mccann_sdonsim_2024,jara_dream-gym_2024}.}
    \label{fig:lit_barchart}
\end{figure}

\subsection{Survey methodology}
\label{sec:survey-methodology}

%we searched the Google Scholar database using the following search terms: "'REINFORCEMENT LEARNING' AND 'NETWORK' AND ('OPTICAL' OR 'WAVELENGTH' OR 'SPECTRUM')". Papers were limited to English-language peer-reviewed articles. We filtered the search results by inspection of paper titles and abstracts, and manually added any related works from our own citation database that were missing from the initial search. A review of the collected papers then created the final set of 97.

To provide an overview of research progress on RL applied to DRA problems in optical networks, we searched to gather all relevant research papers. We performed a manual review of results from citation databases to create the final set of 97 peer-reviewed papers. Figure \ref{fig:lit_barchart} shows the count of papers by publication year. The papers are grouped in 4 categories: 'RWA', 'RSA', 'RMSA', and 'Other'. We use this set of papers for our analysis of benchmarking practices in the field, which we present in the next section with further commentary on Figure \ref{fig:lit_barchart}. 

%the 'Other' category of figure \ref{fig:lit_barchart}, which includes papers on RL applied to: traffic grooming \cite{tanaka_reinforcement-learning-based_2022,zhang_admire_2023,tanaka_adaptive_2023,tanaka_reinforcement-learning-based_2024}, defragmentation \cite{etezadi_deepdefrag_2022,fan_blocking-driven_2023,johari_drl-assisted_2023,etezadi_deep_2023}, survivability or service restoration \cite{zhao_reinforcement-learning-based_2019,luo_leveraging_2019,zhao_service_2021,hernandez-chulde_evaluation_2022,zhao_rsa_2022,luo_survivable_2022,jiao_reliability-oriented_2022}, multicast provisioning \cite{garcia_multicast_2003,tian_reconfiguring_2021,li_tabdeep_2024}, simulation training environments \cite{natalino_optical_2020,natalino_optical_2024,doherty_xlron_2024,mccann_sdonsim_2024}, and other problems such as transceiver parameter optimization \cite{weixer_reinforcement_2020,koch_reinforcement_2022,koch_high-generalizability_2022} or launch power optimization \cite{tse_reinforcement_2024}

\subsection{Review of benchmarking practices}
\label{sec:benchmarking_practices}

%To understand progress on RL applied to DRA problems in optical networks, it is essential to have standard benchmarks. Benchmarks refer to both the problem under investigation and the quality of the solution. In machine learning research, performance benchmarks on MNIST \cite{li_deng_mnist_2012} and ImageNet \cite{krizhevsky_imagenet_2012} have been vital to the astounding progress in that field. 

In optical networks research, the first benchmark for RL was established by DeepRMSA \cite{chen_deeprmsa_2019} (discussed in detail in Section \ref{sec:repro}). DeepRMSA was the first RL approach to achieve lower service blocking probability than KSP-FF, or any heuristic that considers multiple candidate paths. As a result of this breakthrough performance, and its open source codebase, the problem definition from DeepRMSA (topologies, traffic model, modulation format reach, FSU per link, etc.) became a de facto standard. Follow-up works used identical or similar problem definitions and compared to DeepRMSA on their problem \cite{xu_spectrum_2021,quang_magc-rsa_2022,errea_deep_2023,tang_heuristic_2022,xu_deep_2022,cheng_ptrnet-rsa_2024,yan_drl-based_2024,zhou_opti-deeproute_2024}. Arguably, comparing to DeepRMSA has become standard benchmarking practice.

%Although this comparison to the "state of the art" appears to be good practice, it is misguided. The performance of RL agents is highly sensitive to algorithmic details and hyperparameters \cite{engstrom_implementation_2020}, random seeds, and non-deterministic factors \cite{nagarajan_impact_2018}. Guidelines have been established for the reliable comparison of competing RL approaches \cite{henderson_deep_2019}. The comparisons to DeepRMSA in follow-up works have not followed these guidelines. For example, hyperparameters such as learning rate and discount factor should be tuned when applying an algorithm to a new setting. Consequently, the comparisons to DeepRMSA in follow-up works are not fair.

Previous work has called for more rigorous benchmarking practices for research on RL for optical networking \cite{di_cicco_deep_2022}, with recommendations for comparison against other machine learning approaches such as GA and PSO, in addition to estimated bounds on network blocking or throughput. Some studies of RL for resource allocation have restricted themselves to sufficiently small problem sizes and static traffic, to enable comparison to ILP results \cite{liu_waveband_2021,di_cicco_deep_2022,zhao_rsa_2022,momo_ziazet_deep_2022,di_cicco_deepls_2023}. Although this provides a reliable bound, it is not applicable to dynamic traffic.

Benchmarking against standard heuristic algorithms, such as KSP-FF, avoids the complexity of training a competing machine learning approach, performs deterministic allocation, and can scale to large problem instances. However, it is important to choose the best performing heuristic for a particular case of study as a benchmark. Of the papers that benchmark their RL solution to KSP-FF (or other heuristics that consider multiple candidate paths) \cite{chen_deeprmsa_2019,chen_multi-task-learning-based_2021,shi_deep-reinforced_2021,shimoda_deep_2021,xu_spectrum_2021,zhao_reinforced_2021,zhao_service_2021,shimoda_mask_2021,quang_magc-rsa_2022,tu_entropy-based_2022,tang_heuristic_2022,xu_deep_2022,cheng_routing_2022,nevin_techniques_2022,tang_deep_2022,di_cicco_deep_2022,terki_routing_2023,sadeghi_performance_2023,tanaka_adaptive_2023,tang_routing_2023,xu_hierarchical_2023,errea_deep_2023,hernandez-chulde_experimental_2023,cheng_ptrnet-rsa_2024,fan_blocking-driven_2023}, most achieve 20-30\% reduction in service blocking probability compared to their best heuristic. Only 3 papers achieve a reduction greater than this: MaskRSA \cite{shimoda_mask_2021}, PtrNet-RSA \cite{cheng_ptrnet-rsa_2024}, and Terki et al \cite{terki_routing_2022}. Despite these impressive results, we demonstrate in Section \ref{sec:repro_main} that MaskRSA and PtrNet-RSA are beaten by KSP-FF or FF-KSP by considering 50 candidate paths and ordering the paths by number of hops\footnotemark.

\footnotetext{We have not re-created the study of Terki et al. for benchmarking in Section \ref{sec:repro_main} as it is multi-band and out of scope of this work.
We hypothesise that their approach performs strongly because, similar to PtrNet-RSA, it is not limited to selecting from only K paths.}

Benchmarking is further complicated by the fast evolution of optical networking, with novel paradigms such as multi-band \cite{beghelli_approaches_2023} and multi-core \cite{pinto-rios_resource_2023} emerging, and the wide variety of network topologies \cite{matzner_topology_2024} and components that can be considered. The evolution of optical networks research is evidenced by growth in the 'Other' category of Figure \ref{fig:lit_barchart}, which includes papers on RL applied to: traffic grooming \cite{tanaka_reinforcement-learning-based_2022,zhang_admire_2023,tanaka_adaptive_2023,tanaka_reinforcement-learning-based_2024}, defragmentation \cite{etezadi_deepdefrag_2022,fan_blocking-driven_2023,johari_drl-assisted_2023,etezadi_deep_2023}, survivability or service restoration \cite{zhao_reinforcement-learning-based_2019,luo_leveraging_2019,zhao_service_2021,hernandez-chulde_evaluation_2022,zhao_rsa_2022,luo_survivable_2022,jiao_reliability-oriented_2022}, multicast provisioning \cite{garcia_multicast_2003,tian_reconfiguring_2021,li_tabdeep_2024}, %simulation training environments \cite{natalino_optical_2020,natalino_optical_2024,doherty_xlron_2024,mccann_sdonsim_2024}, 
and other problems such as transceiver parameter optimization \cite{weixer_reinforcement_2020,koch_reinforcement_2022,koch_high-generalizability_2022} or launch power optimization \cite{tse_reinforcement_2024}

The establishment of reliable benchmarks is made more difficult by the fragmented software environment for optical network simulations for RL. Several open source toolkits have been introduced to aid researchers and improve productivity, but none has proved sufficiently popular for it to become standard. Optical-rl-gym \cite{natalino_optical_2020} was the first paper to attempt to introduce a new standard library for this task. This was followed by an extension to multi-band environments \cite{morales_multi-band_2021} and in 2024 was further extended to include a more sophisticated physical layer model for lightpath SNR calculations, renamed as the Optical Networking Gym \cite{natalino_optical_2024}. Additionally, MaskRSA provides an open source simulation framework (RSA-RL) \cite{shimoda_mask_2021} and DeepRMSA's codebase is widely used \cite{chen_deeprmsa_nodate}. SDONSim \cite{mccann_sdonsim_2024} and DREAM-ON-GYM \cite{jara_dream-gym_2024} are other recent additions to the landscape of available simulation frameworks that further fragment the available options. 

In summary, progress in applying RL to DRA problems in optical networks has been difficult to quantify due to several factors. First, the lack of standardized benchmarking practices has made it challenging to fairly compare different approaches. Second, while some studies have used ILP solutions as benchmarks, these are limited to small problem sizes and static traffic scenarios, making them impractical for large-scale or dynamic applications. Third, multiple competing simulation frameworks %(Optical-rl-gym, RSA-RL, SDONSim, DREAM-ON-GYM) 
and publications without open source code have made it difficult to ensure consistent testing conditions across different studies. Finally, the rapid evolution of optical networking technology means benchmarks must constantly evolve to remain relevant. 

The lack of reliable benchmarks, and the resulting difficulty in assessing progress in the field, is what motivates our investigations of heuristic benchmarks in Section \ref{sec:heuristic_comparison} and their application to our recreation of previous work in Section \ref{sec:repro_main}.

\subsection{Recommendations for benchmarking best practice}
\label{sec:benchmarking_recommendations}

% Metnion that we do not allow extra cards 

Based on our review of the field, we make the following recommendations for selecting benchmarks and evaluating solutions to DRA problems in optical networks:

\begin{itemize}[itemsep=0pt]
    %\item Use standard problems where possible to facilitate comparison to previous work.
    \item \textbf{Benchmark selection}: Assess multiple benchmarks and select the best available. Tune parameters such as number of candidate paths and path sort criteria (see section 4\ref{sec:path_ordering}) to maximize performance. Prefer deterministic algorithms that can be reproduced without re-training ML components.
    \item \textbf{Statistical rigor}: Perform sufficient trials for statistical significance (minimum 100 blocking events is a rule of thumb) with multiple random seeds. Report mean and a measure of variability of results e.g. standard deviation.
    \item \textbf{Methodological transparency}: Be transparent in your choice of benchmark and its implementation details. Release code to facilitate verification and extension of research findings. Utilize established simulation frameworks where possible to minimize implementation discrepancies across studies, with unit tests to ensure correctness.
\end{itemize}

We implement these recommendations in our analyses in sections \ref{sec:heuristic_comparison} and \ref{sec:repro_main}. We note that these recommendations apply to the evaluation of any solution method for DRA problems, including RL, other ML approaches, or novel heuristic algorithms.

\subsection{Selection of papers for benchmarking}
\label{sec:paper_summaries}

To assess progress in RL for DRA, we select 5 papers to re-benchmark in section \ref{sec:repro_main}. We select these papers primarily because they all compare their results to "DeepRMSA"\footnotemark with similar traffic models and topologies, therefore present the most consistent application of benchmarks in the field. We also select based on their impact, which we assess by qualitative and quantitative criteria. The qualitative criteria are novelty, contribution, and reputation of publication or conference. The quantitative criterion is their blocking performance relative to benchmarks. They are also among the most highly cited papers in the field, as of April 2025.

\footnotetext{We note that the training of RL solutions is highly sensitive to hyperparameters \cite{engstrom_implementation_2020} and non-deterministic factors \cite{nagarajan_impact_2018}, therefore the comparisons that the selected papers make to re-trained DeepRMSA agents may not be robust.}

\begin{enumerate}[itemsep=0pt]
    \item \textbf{DeepRMSA} \cite{chen_deeprmsa_2019} constructs a feature matrix to represent the available paths for the current requests and applies a NN with 5 x 128 hidden units to select from the K-shortest paths with first-fit spectrum allocation. It demonstrates service blocking probability (SBP) reduced by 20\% vs. KSP-FF on the NSFNET and COST239 topologies. DeepRMSA's impact was enhanced by its open source codebase. %published 2019, has over 178 citations.
    \item \textbf{Reward-RMSA} \cite{tang_heuristic_2022} builds on the DeepRMSA framework and changes the reward function to incorporate fragmentation-related information. They report SBP reduced by 32\% vs. multiple heuristics and 55\% vs. DeepRMSA on NSFNET and COST239. %published 2022, has over 22 citations.
    \item \textbf{GCN-RMSA} \cite{xu_deep_2022} is notable as the first work to use advanced NN architectures to improve performance. They use a graph convolutional network (GCN) (including recurrent neural network (RNN) as the path aggregation function) in the policy and value functions, which they claim allows improved feature extraction from the network state. Like DeepRMSA and Reward-RMSA, the policy selects from K paths with first-fit spectrum allocation. They report SBP reduced by up to 30\% vs. multiple heuristics and 18\% vs. DeepRMSA. on NSFNET, COST239, and USNET. %published 2022, has over 36 citations
    \item \textbf{MaskRSA} \cite{shimoda_mask_2021} innovated by selecting from the entire range of available slots on the K paths and using invalid action masking \cite{huang_closer_2022} to increase the efficiency of training. Despite the RSA in the title, the paper does consider distance-dependent modulation format (RMSA). MaskRSA presented improvements over KSP-FF on NSFNET and JPN48 topologies with over an order of magnitude lower SBP, or a 35-45\% increase in the supported traffic in their cases of study. The authors of MaskRSA also contributed to open source by releasing their simulation framework, RSA-RL. %published 2021, has over 28 citations
    \item \textbf{PtrNet-RSA} \cite{cheng_ptrnet-rsa_2024}, published in 2024. It innovates in both the problem setting and its use of pointer-nets \cite{vinyals_pointer_2015}. The pointer-net is used to select the constituent nodes of the target path, thereby removing the restriction of selecting from the pre-calculated K-shortest paths. Invalid action masking is used to allow selection from all available spectral slots. Additionally, the paper considers joint optimization of the mean path SNR and the SBP through its reward function. It demonstrates SBP reduced by over an order of magnitude vs. KSP-FF and their implementation of MaskRSA on NSFNET, COST239, and USNET.
\end{enumerate}

\section{Heuristic algorithm benchmark evaluation}
\label{sec:heuristic_comparison}

To evaluate the results of the selected papers from Section \ref{sec:survey}\ref{sec:paper_summaries}, we must determine the best (lowest blocking probability) heuristic algorithms to use as benchmarks. In this section, we present comparisons of the heuristics listed in Table \ref{tab:heuristics}, evaluated on different traffic loads, topologies, and considering different numbers of candidate paths (K). On the basis of this analysis, we select the benchmarks to apply in section \ref{sec:repro_main}. We also include a discussion of the effect of different sort criteria for candidate paths in Section \ref{sec:path_ordering}, which significantly affects the blocking performance.

We select the algorithms in Table \ref{tab:heuristics} because they are commonly used as benchmarks or have been reported as superior to other heuristics.

\begin{table}[h]
\begin{tabular}{l|l|l}
Heuristic                         & Acronym & Reference                \\ \hline
K-Shortest Paths First-Fit        & KSP-FF  & \cite{zang_review_2000} \\
First-Fit K-Shortest Paths        & FF-KSP  & \cite{vincent_scalable_2019} \\
K-Shortest Paths Best-Fit         & KSP-BF  & \cite{abkenar_best_2016} \\
Best-Fit K-Shortest Paths         & BF-KSP  & \cite{abkenar_best_2016} \\
K-Minimum Entropy First-Fit       & KME-FF  & \cite{wright_minimum-_2015} \\
%K-Minimum Cut First-Fit           & KMC-FF  & \cite{tang_heuristic_2022} \\
%K-Minimum Frag First-Fit & KMF-FF  & \cite{tang_heuristic_2022} \\
K-Congestion Aware First-Fit      & KCA-FF  & CA2 from \cite{savory_congestion_2014}
\end{tabular}
\caption{RMSA heuristics used for benchmarking.}
\label{tab:heuristics}
\end{table}

%We also evaluated the K-Minimum Cut First-Fit (KMC-FF) and K-Minimum Frag First-Fit (KMF-FF) heuristics from Reward-RMSA \cite{tang_heuristic_2022}. These heuristics produce higher blocking probability than others when paths are ordered by number of hops. We exclude these results to improve the clarity of Figures \ref{fig:heur_comp}, \ref{fig:k_traffic}, and \ref{fig:heur_traffic}.

\begin{figure}
  \includegraphics[width=1.01\linewidth]{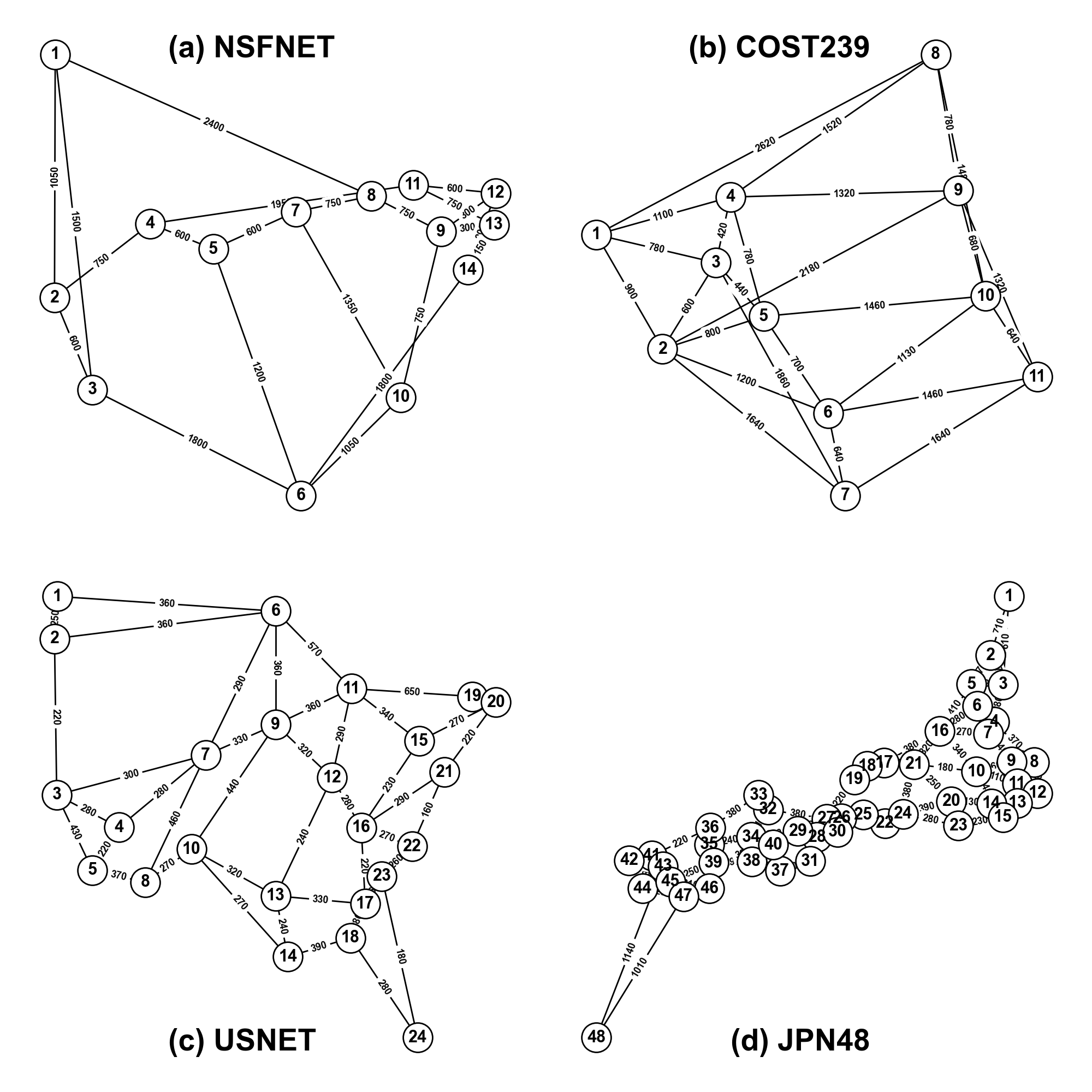}
  \caption{Network topologies used in our case studies from: DeepRMSA, Reward-RMSA, GCN-RMSA, MaskRSA, PtrNet-RSA \cite{chen_deeprmsa_2019} \cite{tang_heuristic_2022} \cite{xu_deep_2022} \cite{shimoda_mask_2021} \cite{cheng_ptrnet-rsa_2024}. We note that the USNET topology differs between GCN-RMSA and PtrNet-RSA. We show the GCN-RMSA version here. PtrNet-RSA also uses a variant of the COST239 topology.} %The COST239 topology shown is from DeepRMSA, the USNET is from PtrNet-RSA.}
  \label{fig:network_plots}
\end{figure}

Figure \ref{fig:network_plots} shows the topologies used in the selected papers. The node and link count for each topology is: NSFNET (14, 22), COST239 (11, 25), USNET (24, 43), JPN48 (48, 82). We use these topologies to analyze the performance of the heuristic algorithms and in our recreation of the papers' problems in section \ref{sec:repro_main}. %Some characteristic features of the topologies are summarised in table \ref{tab:topology_features}. 
We make all topology data available in our open source codebase \cite{michael_doherty_2024_jocn_xlron_2024}.

\subsection{Effect of path ordering}
\label{sec:path_ordering}

All the heuristics in Table \ref{tab:heuristics} select from the available pre-computed paths on the basis of sort criteria. The primary criterion may be a measure of the path congestion (KCA-FF), spectral fragmentation (KME-FF), or length (KSP-FF). In the event of multiple paths with equal value, a default ordering (usually ascending order of length) determines the selected path.

Conventionally, path length is considered as distance in km. However, we find that considering path length as number of hops (with length in km as a secondary sort criterion), significantly improves the performance of the heuristics. This has been observed previously by Baroni \cite{baroni_routing_1998}, who referred to it as Minimum Number of Hops routing (MNH). The intuitive explanation for this is that, if two paths can support the same order of modulation format, the path that comprises fewer links occupies fewer spectral resources. 

We refer to these two orderings as path length in km (\#km) or path length in number of hops (\#hops). Our comparisons of KSP-FF for these two orderings in Section \ref{sec:repro_main} Figure \ref{fig:repro} evidence the reduction in blocking probability from \#hops ordering. In our comparisons of heuristics in the next section, we use \#hops ordering.

\subsection{Simulation setup}

For each heuristic and topology, we carried out three simulation scenarios to investigate the effects of varying traffic loads and values of K on the relative blocking performance of the heuristics. 

\vspace{0.1cm}

\noindent \textbf{Experiment 1} - Increasing K:
\par \noindent \textbf{Aim}: Investigate relative performance of heuristics with increasing K. \textbf{Method}: Record service blocking probability (SBP) for each heuristic at values of K ranging from 2 to 26 at fixed traffic load. We arbitrarily select the traffic load for each topology so that the heuristics give a SBP of  $\sim$1\%.

\vspace{0.1cm}

\noindent \textbf{Experiment 2} - Increasing K at high to low traffic: 
\par \noindent \textbf{Aim}: Investigate the effect of increasing K at different traffic loads. \textbf{Method}: Record SBP at K ranging from 2 to 40 for a range of traffic loads. We select the traffic loads for each topology such that they result in 10$^{-5}$ to 10$^{-1}$ SBP. To simplify the analysis and plots, we only present results for KSP-FF.

\vspace{0.1cm}

\noindent \textbf{Experiment 3} - Increasing traffic load at K=50: 
\par \noindent \textbf{Aim}: Using the findings from Experiments 1 and 2, determine the lowest-blocking heuristic with optimized K-value across traffic loads. \textbf{Method}: Record SBP for high K (K=50) at varying traffic loads. We select the traffic loads for each topology such that they result in a range of SBP (10$^{-5}$ to 10$^{-1}$). This experiment provides evidence on which heuristic is the best overall for each topology.

\vspace{0.1cm}

For each experiment and heuristic, data was collected from 3000 independent trials with unique random seeds. The SBP was calculated after 10,000 connection requests, with the mean and standard deviation calculated across trials. Each data point in Figure \ref{fig:heuristic_combined} therefore shows summary statistics from 30 million connection requests, which gives high confidence in our results.

We considered dynamic traffic with fixed mean service holding time at 10 units. We considered the same traffic model and other settings as DeepRMSA\footnotemark: uniform traffic probability between each node pair, Poissonian arrival and departure statistics, uniform random selection of data rate from 25 to 100Gbps in 1Gbps intervals, and distance-dependent modulation formats from BPSK, QPSK, 8QAM, 16QAM, and maximum transmission distances of 10,000km, 2500km, 1250km, 625km, respectively,  We consider topologies with dual fibre links (one for each direction of propagation), 12.5GHz FSU width, and 100 FSU per fibre.

\footnotetext{We consider the DeepRMSA problem settings in these experiments because it is used by most of the papers presented in Section \ref{sec:repro_main}.} %For the same reason, we report our results in terms of SBP, despite our recommendation to use BBP in Section \ref{sec:recommendations}.}

\subsection{Results and discussion}

\begin{figure*}[ht]
    \centering
    \includegraphics[width=\textwidth]{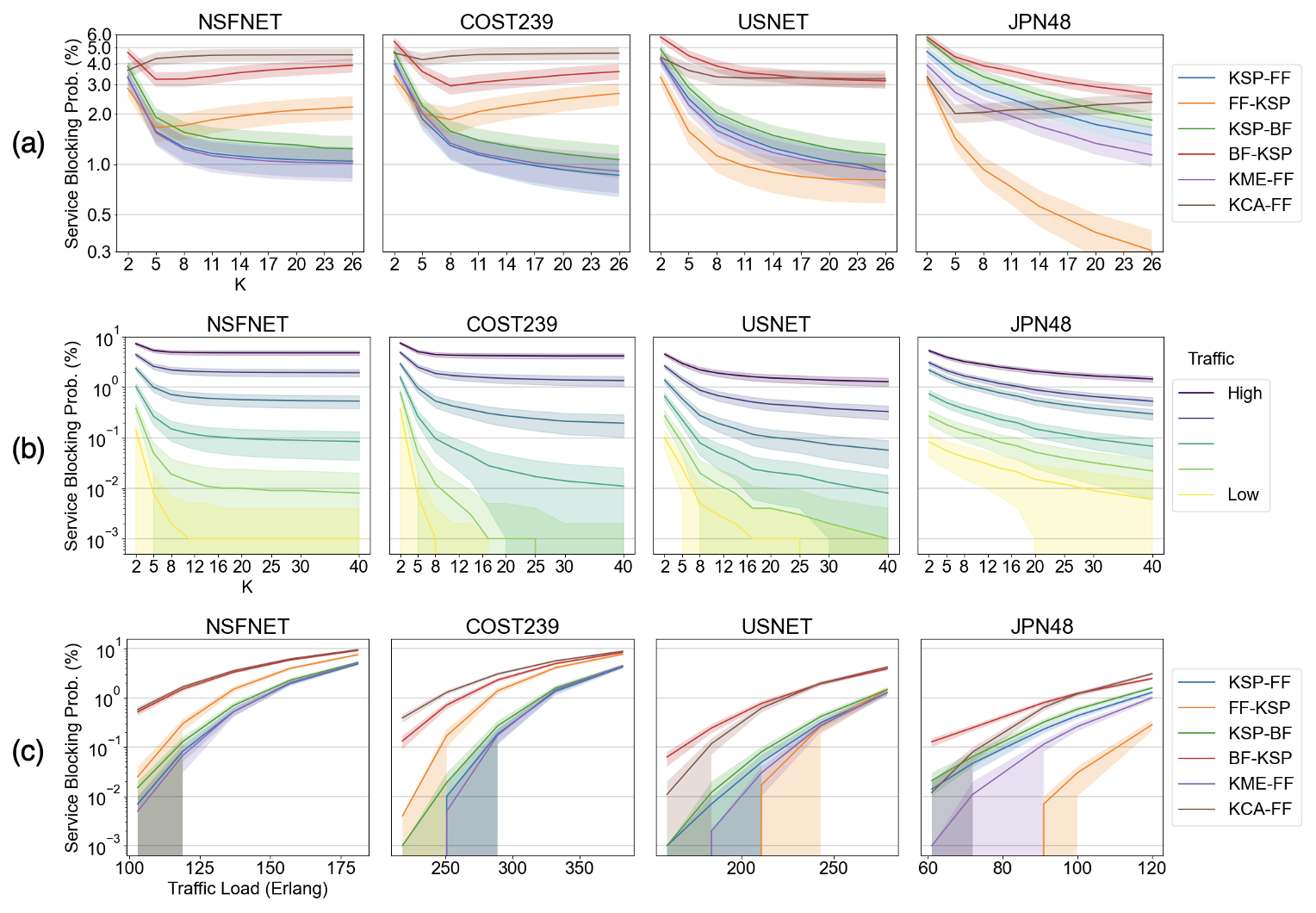}
    \caption{Comparison of heuristic algorithms. (a) Service blocking probability (SBP) at fixed traffic and varying numbers of candidate paths (K). (b) SBP for KSP-FF at varying traffic loads and K=2 to K=40. (c) SBP at varying traffic load for K=50. The mean and standard deviation (shaded area) are calculated from 3000 trials of 10,000 traffic requests per data point. KSP-FF or FF-KSP with K=50 are found to give the lowest blocking.}
    \label{fig:heuristic_combined}
\end{figure*}

\textbf{Experiment 1} results in Figure \ref{fig:heuristic_combined}(a) show different outcomes for smaller networks (NSFNET and COST239) and larger networks (USNET and JPN48). For NSFNET and COST239, KSP-FF and KME-FF are approximately equal and give the lowest blocking. Their blocking decreases to a minimum for approximately K=23 and above for NSFNET and continues to decrease for K>26 for COST239.

For USNET and JPN48, FF-KSP is clearly the best heuristic, with blocking reduced by half for JPN48. Blocking from FF-KSP decreases with K until K=26 for USNET and continues dropping sharply for K>26 for JPN48. For USNET, KSP-FF and KME-FF become competitive with FF-KSP at large K. It can be argued that FF-KSP performs better in networks with higher numbers of nodes and links where there are many viable paths between source and destination, and dense packing of utilized wavelengths increases in relative importance to path selection.

The results from Experiment 1 indicate that KSP-FF and FF-KSP generally give the lowest blocking, depending on the network topology, and blocking decreases monotonically with increasing K. This experiment looked at a moderately high traffic load ($\sim$1\% SBP), therefore experiment 2 investigates if the effect of increasing K holds at different traffic loads. 

\textbf{Experiment 2} results in Figure \ref{fig:heuristic_combined}(b) show that, regardless of the traffic load, increasing K decreases the SBP, until SBP reaches a minimum and increasing K does not decrease SBP further. Across all topologies and traffic loads tested in our experiments, we found that SBP does not decrease significantly for K>50. For very high traffic (approximately equivalent to incremental loading), the value of K beyond which SBP does not continue to decrease can be much lower.

\textbf{Experiment 3} results in Figure \ref{fig:heuristic_combined}(c) show the variation of SBP with traffic load for each heuristic with K=50. We verified that at least 50 unique paths are possible for every node pair on our investigated topologies. We select K=50 on the basis of experiments 1 and 2. These results confirm the initial findings from Experiment 1 - that KSP-FF and KME-FF are the lowest blocking for NSFNET and COST239\footnotemark, while FF-KSP is better for USNET and JPN48, with an order of magnitude lower blocking probability on JPN48 compared to the next best heuristic.

\footnotetext{Although Figure \ref{fig:heuristic_combined}(c) shows KME-FF gives slightly lower blocking than KSP-FF at lower traffic, we prefer KSP-FF for benchmarking purposes because of its widespread use and its greater simplicity.}

In summary, we highlight the generally strong performance of the KSP-FF and FF-KSP heuristics. We find that increasing the number of candidate paths decreases the blocking probability, as does the ordering of candidate paths. We find \#hops ordering is superior to \#km for reduced blocking probability, as evidenced in Section \ref{sec:repro_main} Figure \ref{fig:repro}.

We point out that the list of heuristics we evaluate is not exhaustive and superior algorithms may exist. We therefore encourage thorough analysis to determine the strongest heuristic benchmark for a particular problem, as we have exemplified here. However, for the purposes of this study and our comparisons to previous work in Section \ref{sec:repro_main}, we find that KSP-FF with K=50 and \#hops ordering demonstrates lower blocking probability than previous RL approaches.

\section{Benchmarking of previous work}
\label{sec:repro_main}

As discussed in our literature review (Section \ref{sec:survey}), it is difficult to assess progress in the field due to several factors, particularly the diversity of problem definitions and use of weak benchmarks. To address this, we exactly recreate the problem settings from five influential papers from the literature, and apply the best-performing heuristics from Section \ref{sec:heuristic_comparison} in each case.

In this section, we first provide analysis of holding time truncation, an implementation detail present in the DeepRMSA codebase that significantly affects the blocking probability. We then present the results of our reproductions of the selected papers and compare to the heuristics, which have significantly lower blocking probability all of the published RL solutions.

We have corresponded with the authors of the selected papers to clarify details of their implementation and ensure that our recreations exactly match all the relevant details of their problems. The table of Appendix A provides numerical comparisons of the results of KSP-FF from the papers and our recreation of their problems, which show good agreement within one standard error. We point out that we do not reproduce the training of the published RL results. We choose not to reproduce training because of insufficient training details and the widely documented difficulties in reproduction of RL training due to sensitivity to hyperparameters and random seeds \cite{henderson_deep_2019,engstrom_implementation_2020}. Extracting RL results from published papers gives a more reliable and fair comparison.

 We use our high-performance simulation framework, \mbox{XLRON} \cite{doherty_xlron_2023}, for all experiments. It has demonstrated 10x faster execution on CPU and over 1000x faster when parallelized on GPU compared to optical-rl-gym \cite{doherty_xlron_2024}. This is possible due to its array-based data model and use of the JAX numerical array computing framework, that enables just-in-time compilation to accelerator hardware. It also offers a complete suite of unit tests for core functionality, making it reliable, and includes features to reproduce the problem settings of the selected papers. We use it for these reasons and for its simple command-line interface, which facilitates experiment automation and reproducibility.

\subsection{Holding time truncation}
\label{sec:holding_time}

In dynamic traffic simulations, the service holding time and time until the next arrival of a service request are modeled as exponentially distributed random variables, which is consistent with the assumption of Poisson arrival processes. Random sampling from these exponential distributions is used to generate times for each service request in the simulation.

DeepRMSA, Reward-RMSA, and GCN-RMSA use the same original DeepRMSA codebase as the basis for their experiments. This codebase includes a significant detail: the service holding time is resampled if the resulting value is more than twice the mean of the distribution. We refer to this detail as holding time truncation.  In order to recreate the problems from these papers, we analyze the effect of holding time truncation.

\subsubsection{Experiment setup}
To understand the effect of truncation on the traffic statistics, we define an exponential distribution with unit mean. We take $10^{6}$ samples from the distribution, both with and without truncation, and calculate the mean of the resulting sample populations in both cases.

\subsubsection{Results and discussion}
\begin{figure}
    \centering
    \includegraphics[width=0.9\linewidth]{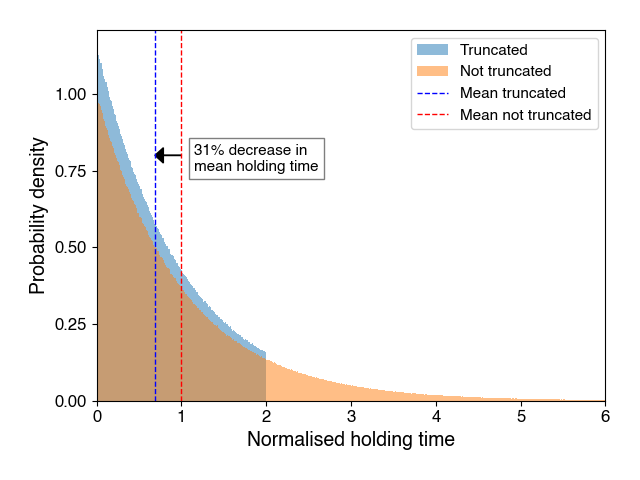}
    \caption{Histogram of service holding holding times. The truncated distribution resamples the holding time when the sampled value exceeds $2*$mean. This reduces the mean holding time by 31\% compared to the standard exponential distribution.}
    \label{fig:truncation}
\end{figure}

Figure \ref{fig:truncation} compares histograms of service holding times with and without truncation. The y-axis shows the probability density, which is normalized so the area of each histogram gives unit probability. The truncated case shows a cutoff at twice the mean holding time. The vertical lines indicate the mean for each case.

Holding time truncation reduces the mean by approximately 31\%. This results in 31\% lower traffic load. Therefore, papers that use the DeepRMSA codebase (including DeepRMSA, Reward-RMSA, and GCN-RMSA) evaluate their solutions at traffic loads 31\% lower than reported. This detail is not made explicit in the published papers. This finding highlights the challenges in making fair comparisons between papers, and the need for transparency in research code.

\subsection{Benchmarking of published results}
\label{sec:repro}
Our analysis of the best performing heuristics, of holding time truncation, and our correspondence with the authors enables us to benchmark the published results from the five selected influential papers. The aim of this comparison is to determine if any of the published RL solutions achieve lower SBP than the heuristics. %In order to do so, we recreate their simulations exactly and apply the KSP-FF heuristic algorithm with K=50 and paths ordered by number of hops in each case, based on our experiments from section \ref{sec:heuristic_comparison}.

\subsubsection{Experiment methodology}

We recreate the problems from each selected paper in our own simulation framework \cite{doherty_xlron_2024}. We match the topologies (NSFNET, COST239, JPN48, USNET), mean service arrival rates, mean service holding times, data-rate or bandwidth request distributions, and uniform traffic matrices. We use the same measurement methodology as described in the respective papers to reproduce results, which is 3000 request warm-up period (to allow the network blocking probability to reach steady-state after the 'initial transient' \cite{white_problem_2009}) followed by 10,000 requests. The SBP is calculated at the end of the episode. We run 10 independent episodes at each traffic load per problem and calculate the mean and standard deviation across trials.

We extract published results for KSP-FF and RL solutions from the papers, using textual values where available otherwise reading from charts. All published results report only a single data point for each traffic value, without uncertainty estimates. %We provide all of the extracted data and our results in Appendix \ref{appendix:A}.

We check that our results for KSP-FF with K=5 (green line in Figure \ref{fig:repro}) match the published results for KSP-FF (blue line in Figure \ref{fig:repro}) within two standard deviations to ensure faithful reproduction. This comparison gives us a high degree of confidence that we have exactly recreated each problem setting.

\subsubsection{Results and discussion}

\begin{figure*}[ht]
  \includegraphics[width=1.01\textwidth]{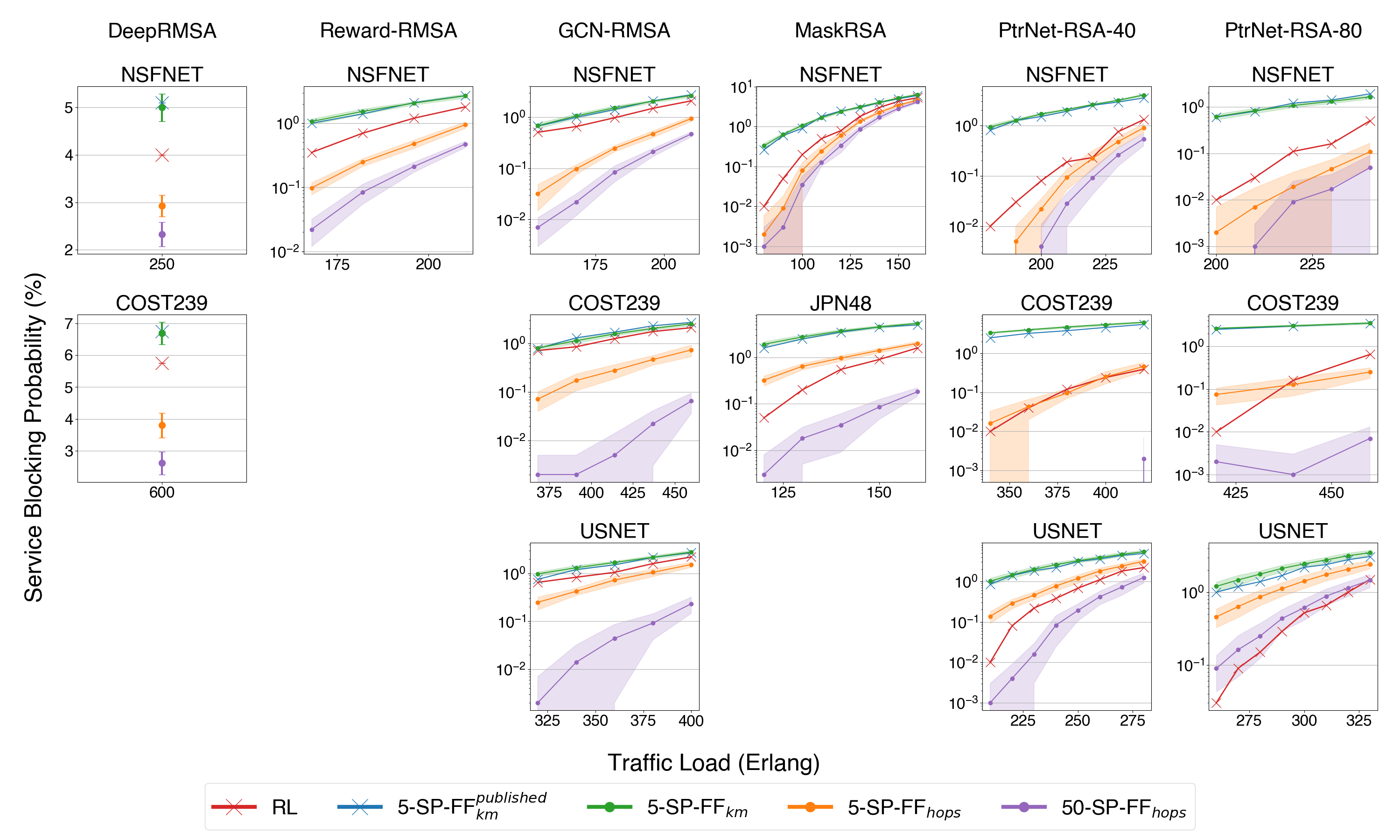}
  \caption{Mean SBP against traffic load. Each column is a publication and each subplot is for a topology. Error bars and shaded areas show standard deviations. %Data for $RL$ and 5-SP-FF$_{published}$ are extracted from the publications. Close agreement between 5-SP-FF$_{published}$ and  5-SP-FF$_{ours}$ show that we accurately reproduce each case of study.
  \mbox{50-SP-FF$_{hops}$} exceeds or matches the $RL$ performance for each case.}
  \label{fig:repro}
\end{figure*}

Figure \ref{fig:repro} shows our reproduction of results from the selected papers, with SBP against traffic load in Erlangs in each subplot. The plots are organized by paper (columns) and topology (rows). PtrNet-RSA has two columns reflecting its two test cases: networks with 40 FSU per link and 1 FSU requests, and networks with 80 FSU per link and 1-4 FSU requests. PtrNet-RSA only considers fixed-bandwidth requests (no distance-dependent modulation format). MaskRSA and PtrNet-RSA only consider single fibre links (counter-propagating channels) , whereas the other cases consider dual fibre links (one fibre for each direction of propagation), which increases their capacity.

Each plot contains 5 datasets:
\begin{itemize}[itemsep=0pt]
\item[] \textbf{RL}: Published results for the RL approach
\item[] \textbf{5-SP-FF$^{published}_{km}$}: Published results for KSP-FF (K=5) with paths ordered by \#km
\item[] \textbf{5-SP-FF$_{km}$}: Our results for KSP-FF (K=5) with paths ordered by \#km
\item[] \textbf{5-SP-FF$_{hops}$}: Our results for KSP-FF (K=5) with paths ordered by \#hops
\item[] \textbf{50-SP-FF$_{hops}$}: Our results for KSP-FF (K=50) with paths ordered by \#hops
\end{itemize}

Points show mean values, shaded areas indicate standard deviation and lines interpolate between points. The DeepRMSA paper provides data for only one traffic load per topology. The excellent agreement between 5-SP-FF$^{published}_{km}$ and 5-SP-FF$_{km}$ in all cases confirms that our framework accurately reproduces the published scenarios. %We can therefore make comparisons between our results on these problem settings and the published RL results with high confidence.

From Figure \ref{fig:repro}, we highlight the comparisons of 'RL' (red) with 5-SP-FF$_{hops}$ (orange), and 50-SP-FF$_{hops}$ (purple). 5-SP-FF$_{hops}$ reduces the blocking probability by up to an order of magnitude compared to RL in all cases for NSFNET, 4/5 cases for COST239 and 1/3 cases for USNET. This shows that ordering paths by \#hops is sufficient to beat the RL results in these cases.

For larger topologies, considering more candidate paths (K>5) improves the heuristic performance significantly, often by over an order of magnitude. As shown by Figure \ref{fig:repro}, 50-SP-FF$_{hops}$ gives the lowest SBP of all approaches in all cases, except PtrNet-RSA-80 USNET (bottom right).

For PtrNet-RSA-80 USNET, we consider it plausible that the pointer-net architecture is a contributing factor to the strong performance, as it is not limited to selecting from a pre-defined set of paths. However, as the published results in this case fall within one standard deviations of the mean for 50-SP-FF$_{hops}$, the result could be spurious. This highlights the need for summary statistics and confidence intervals from multiple trials to be included with published results.

In summary, the results show that making minor changes (ordering paths by \#hops and considering more paths)  to simple heuristic algorithms is sufficient to achieve lower blocking probability than the sophisticated RL solutions that have been published.

We highlight that this analysis, and the selected papers, focus on SBP as the optimization objective. In realistic scenarios, network blocking or throughput must be balanced with other metrics such as latency and total cost of operation from transceiver launch power, amplifiers, and other network elements. Future research should therefore focus on problems that take a holistic approach to network operations optimization with multiple objectives \cite{nallaperuma_interpreting_2023}, and incorporate sophisticated models of all physical layer effects for improved accuracy \cite{curri_gnpy_2022,buglia_closed-form_2023}.

All data shown in Figure \ref{fig:repro} is provided in tabular form in Appendix A.

\section{Network blocking bounds}
\label{sec:bounds}

We have demonstrated in Section \ref{sec:repro_main} that many influential works on RL for DRA problems in optical networks have failed to improve on a simple heuristic algorithm. The extent to which it is possible to reduce the blocking probability, and increase supported traffic, is an important motivating factor in any future research into this topic. 

To understand the limits of blocking probability, we derive empirical lower bounds. By comparing these lower bounds to the performance of our best solution for a target SBP, we can estimate the additional traffic load that can be supported and, therefore, the maximum benefit from applying an intelligent resource allocation method such as RL. %We term this additional capacity the "optimality gap", as we consider our lower bound to be approximately optimal.

As discussed in section \ref{sec:background}, DRA problems in optical networks that require RSA are subject to three constraints: spectrum continuity, spectrum contiguity, and no reconfiguration. By relaxing any of these constraints, the optimal or near-optimal solution of the relaxed problem is a bound on the solution of the full problem. The cut-sets bound method of Cruzado et al \cite{cruzado_effective_2023,cruzado_capacity-bound_2024} relaxes the spectrum continuity constraint and uses insights from the min-cut max-flow theorem to estimate a lower bound SBP. We instead relax the constraint on reconfiguring already-established connections, a process known as defragmentation.

We couple this defragmentation with resource prioritization: sorting the active connection requests by their required resources and allocating them sequentially. The sorting of active requests in descending order of required resources was found to improve the achievable capacity to optimal or near-optimal by Baroni \cite{baroni_routing_1998} in static RWA and later Beghelli \cite{beghelli_resource_2006} for dynamic RWA, a method they refer to as 'reconfigurable routing'. Since our problem settings are elastic optical networks, we prefer the term defragmentation. The intuition behind this approach is to allocate requests with longer paths and higher spectral requirements first so that requests with lower resource requirements may be squeezed into remaining spectral gaps later.

\subsection*{Resource-Prioritized Defragmentation}
\label{sec:bounds}

\begin{algorithm}
\caption{Resource-Prioritized Defragmentation Blocking Bound Estimation}
\begin{algorithmic}[1]
\Require Network topology $G$, Set of requests $\mathcal{R}$, Frequency slots per link $F$
\Ensure Blocking probability $P_b$
\State $N \gets \textsc{InitializeNetwork}(G, F)$ \Comment{Initialize network state with empty spectrum slots}
\State $\textit{blocked} \gets \texttt{false}$
\State $\textit{blocked\_requests} \gets 0 $
\For{$t \gets 1$ to $|\mathcal{R}|$}
    \State $N \gets \textsc{RemoveExpiredRequests}(N, t)$
    \State $r_t \gets \textit{current request from } \mathcal{R}$
    \State $N$, $\textit{blocked} \gets \textsc{AllocateRequest}(N, r_t)$
    
    \If{$\textit{blocked}$}
        \State $\textit{active\_requests} \gets \textsc{GetActiveRequests}(\mathcal{R}, t)$
        \State $\textit{sorted\_requests} \gets \textsc{SortByResource}(\textit{active\_requests})$
        \State $N_{temp} \gets \textsc{InitializeNetwork}(G, F)$
        \State $\textit{blocked} \gets \texttt{false}$
        
        \For{$r \in \textit{sorted\_requests}$}
            \State $N$, $\textit{blocked} \gets \textsc{AllocateRequest}(N_{temp}, r)$
            \If{$\textit{blocked}$}
                \State \textbf{break}
            \EndIf
        \EndFor
        
        \If{not $\textit{blocked}$}
            \State $N \gets N_{temp}$
        \Else
            \State $\textit{blocked\_requests} \gets \textit{blocked\_requests} + 1$
        \EndIf
    \EndIf
\EndFor

\State \Return $\frac{\textit{blocked\_requests}}{|\mathcal{R}|}$
\end{algorithmic}
\label{algo:defrag}
\end{algorithm}

The resource-prioritized defragmentation algorithm is outlined in Algorithm \ref{algo:defrag}. It utilizes four key subroutines:

\begin{itemize}
    \item \textsc{RemoveExpiredRequests}($N$, $t$) maintains network state by removing connections that have expeired. For current time $t$, and request with arrival time $t_{\text{arrival}}$ and holding time $t_{\text{holding}}$, the expiry condition is defined as: $t_{\text{arrival}} + t_{\text{holding}} < t$.
    
    \item \textsc{AllocateRequest}($N$, $request$) establishes a new connection subject to continuity and contiguity constraints, and returns the updated network state and a boolean to indicate if the connection was blocked. We use the KSP-FF or FF-KSP algorithm with K=50. We select the algorithm that produces the lowest SBP for the problem instance.
    
    \item \textsc{GetActiveRequests}($\mathcal{R}$, $t$) identifies requests where $t_{\text{arrival}} \leq t < t_{\text{arrival}} + t_{\text{holding}}$, determining which connections require reallocation during defragmentation.
    
    \item \textsc{SortByResource}($requests$) orders active requests by required resources (product of required spectral slots and hops of shortest path), prioritizing larger requests during reallocation to maximize the probability of finding viable configurations.
    
\end{itemize}

A shortcoming of our method of blocking bound estimation is its reliance on the internal \textsc{AllocateRequest} heuristic. To have confidence that the solution presents a true bound, the allocation method must be as close to optimal as possible. We therefore evaluate multiple heuristics for each case, as shown in Section \ref{sec:heuristic_comparison}, and select the one with lowest SBP. We find the best performing heuristic is KSP-FF$_{hops}$ with K=50 for most cases, except MaskRSA JPN48 which is FF-KSP.

%Combined with resource prioritization (\textsc{SortByResource}) we assume the results are near-optimal, based on results from Baroni \cite{baroni_routing_1998}. 
An advantage of our method compared to cut-sets analysis is it computes an allocation that is guaranteed to be physically possible, as it relaxes the 'No Reconfiguration' constraint instead of the physical spectrum continuity constraint. Relaxing the 'No Reconfiguration' constraint makes Algorithm \ref{algo:defrag} omniscient (it has complete knowledge of requests to be allocated) rather than a strictly on-line algorithm, according to definitions from Awerbuch et al \cite{awerbuch_throughput-competitive_1993}. This gives Algorithm \ref{algo:defrag} a fundamental competitive advantage over on-line algorithms like KSP-FF/FF-KSP, therefore it can be considered a lower bound estimator of blocking probability. % This disparity in information may however mean it is too much of an upper-bound estimate
%Overall, resource-prioritized defragmentation bounds can be considered complimentary to cut-sets bounds due to the difference in constraint relaxation.

We note that our algorithm is general and can be applied to any DRA problem in optical networks by using a strong heuristic for \textsc{AllocateRequest} and defining the resource-based sort criteria appropriately.

\begin{figure*}[t]
  \includegraphics[width=1.01\textwidth]{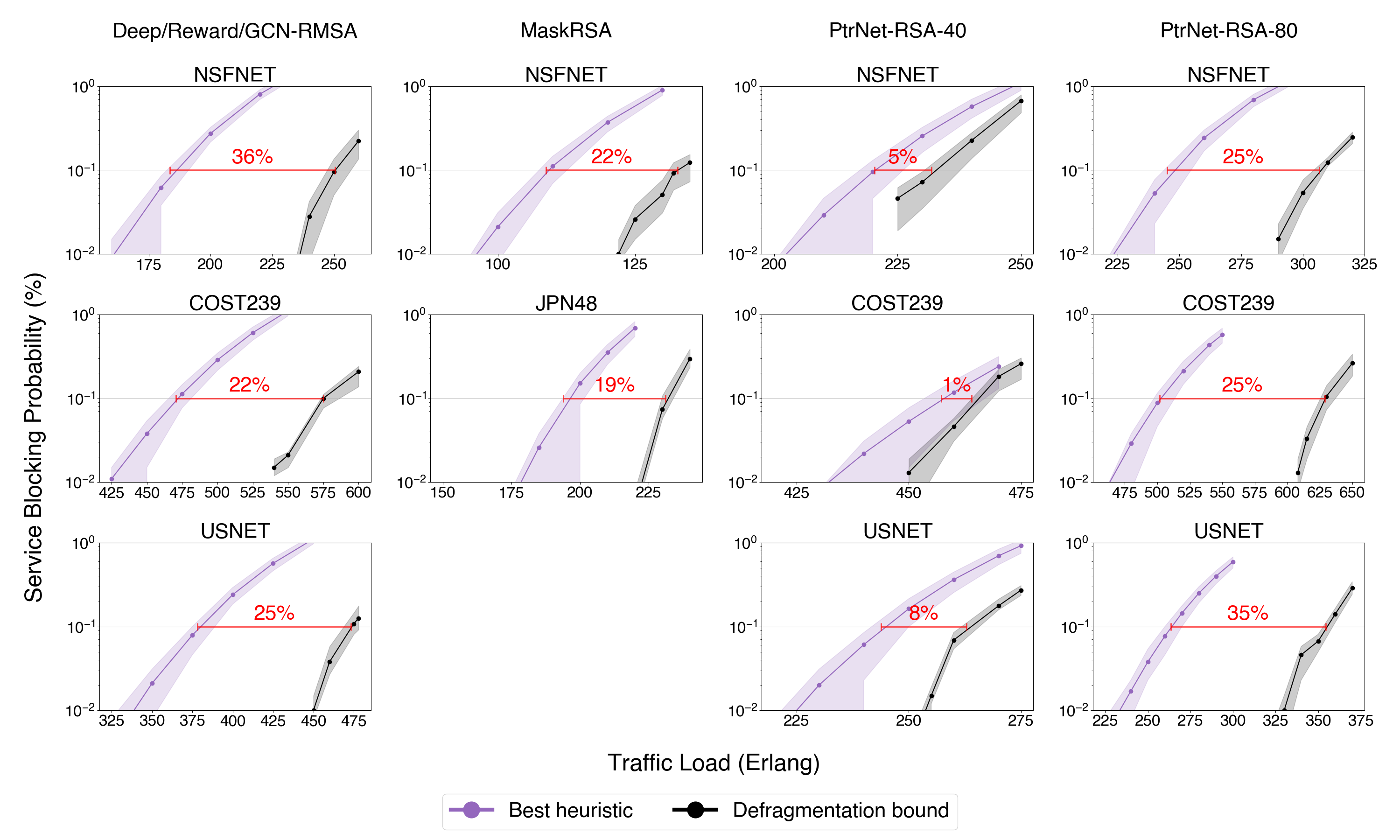}
  \caption{Mean SBP against traffic load for the lowest-blocking heuristic in each case (KSP-FF or FF-KSP with K=50) and the estimated bound from Algorithm \ref{algo:defrag}. Each column is a publication and each subplot is for a topology. Shaded areas show standard deviations. Red lines and text indicate relative increase in supported traffic at 0.1\% SBP from heuristic to bound.}
  \label{fig:bounds}
\end{figure*}

\subsection{Experiment setup}

For each problem from the five selected papers, we run the best performing heuristic for a range of traffic loads that result in SBP from 0.01\% to 1\%. For the lowest-blocking heuristic and for Algorithm \ref{algo:defrag}, we run 10 episodes of 10,000 requests with unique random seeds and calculate the mean and standard deviation of SBP across episodes. We calculate the mean and standard deviation SBP across episodes in each case. 

We compare the resulting SBP from the best heuristic and from algorithm \ref{algo:defrag}. We seek to estimate the additional network capacity that can be achieved at 0.1\% SBP for each case of study from the five selected papers. We select 0.1\% SBP to align with previous studies of network throughput estimation by Cruzado et al \cite{cruzado_effective_2023,cruzado_capacity-bound_2024}.

\subsection{Results and discussion}

Similar to Figure \ref{fig:repro}, each subplot in Figure \ref{fig:bounds} represents a different problem instance. DeepRMSA, Reward-RMSA, and GCN-RMSA are combined into a single set of plots since they use identical topologies and traffic models. The purple lines show the best performing heuristic in each case (KSP-FF with K=50, or FF-KSP for JPN48), with paths sorted in ascending order of number of hops. The grey lines show the resource-prioritized defragmentation bounds. At 0.1\% SBP, we compare the network traffic loads that can be supported in each case, with the difference highlighted by a red horizontal line. The relative increase in network capacity is calculated as the difference between the upper bound traffic load and the heuristic traffic load, as a percentage of the heuristic load.

PtrNet-RSA-40 shows differences of 5\%, 1\%, and 8\% across its three test cases. These relatively low values are due to the fixed width requests size of 1 FSU used in this case, which makes it equivalent to RWA and reduces the impact of fragmentation compared to RSA/RMSA.

For the Deep/Reward/GCN-RMSA, MaskRSA and PtrNet-RSA-80 cases, the difference between the supported traffic in the heuristic case and the upper bound ranges from 19\% (MaskRSA JPN48) to 36\% (Deep/Reward/GCN-RMSA NSFNET). These results show larger but comparable optimality gaps to those from the cut-sets method of Cruzado et al. \cite{cruzado_capacity-bound_2024}, who found gaps of 5\% to 16\% in their cases of study. This shows that defragmentation can unlock significant network capacity, but it is unknown theoretically how close an intelligent online allocation method, such as RL, can come to this bound. This will be the subject of future research.

\section{Conclusion}
\label{sec:conclusion}

Our review of the field of RL applied to DRA problems in optical networks shows that it has been the subject of significant research interest, with almost 100 peer-reviewed papers published so far. Technical innovations from ML research, such as invalid action masking \cite{shimoda_mask_2021,nevin_techniques_2022,cheng_ptrnet-rsa_2024} and GNNs \cite{xu_deep_2022,li_gnn-based_2022,xiong_graph_2024}, have been applied to the problem area and have demonstrated incremental improvements in network blocking.

However, the field has suffered from a lack of standardization in problems, selective application of benchmark algorithms, and poor practices for reproducibility. We have addressed these problems by assessing a range of heuristic algorithms, optimizing their path ordering and number of candidate paths, and applying them to the problem settings from five influential papers on RL for DRA. We use our simulation framework for this work, which enabled recreation of diverse problem settings and fast computation.

From our assessment and optimization of heuristic algorithms, we determine that KSP-FF or FF-KSP with K=50 are the best of those we evaluated. We highlight the result that ordering the candidate paths by number of hops gives significantly lower blocking probability than ordering by distance. These recommended benchmarks can be applied to future studies of RL or other solution methods.

 %Our quantification of the effects on mean path length of reordering paths in this way show that the mean length is increased by around 20\% in most cases, which is likely an acceptable trade-off in exchange for greater network capacity. %We also quantify the effect of network warm-up, and make recommendations that the warm-up period should be at least 7x the target traffic load to allow stabilization before training or evaluating on dynamic network traffic.

Our most significant findings are in the benchmarking of previous RL results. By extracting the published results of RL from the selected papers, and comparing to the best heuristic benchmarks on recreated problem settings, we show that simple heuristics exceed or match the RL results in all cases, often with over an order of magnitude lower SBP. This shows the relative performance of previous RL solutions on these problems has been overestimated due to weak benchmarks, and highlights the need for more rigorous standards of evaluation on these problems to avoid trivial results. These standards also apply to other non-RL resource allocation algorithms.

Finally, to ascertain the practical value of pursuing further research into optimized DRA, we provide the Resource-Prioritized Defragmentation method of estimating the lower bound network blocking probability. Compared to the best heuristics available for each case, this method estimates the upper bound additional dynamic traffic load that can be supported on flex-grid networks is approximately 19\% to 36\%. These results suggest there is room for improvement over the best benchmarks, which may motivate further research into DRA with RL or other methods. Alternatively, research into network optimization with RL could focus on other objectives for which there are not yet good heuristic solutions.

%Considering that IP traffic is projected to increase at approximately 20\% annually until 2029 \cite{ericsson_ab_ericsson_2024}, it is an open question if the potential percentage gains in network throughput are worth pursuing with RL, or if research into network optimization with RL should focus on other objectives for which there are not yet good heuristic solutions.

% Our method of capacity bound estimation is simple and scalable, and can be used by other researchers to estimate the quality of their solution. It is also adaptable to other DRA problems

% We show 2 important qunatitative results. 1) Heuristics give lower service blocking probability than Rl for all published works. 2)

\section*{Acknowledgments}
This work was supported by the Engineering and Physical Sciences Research Council (EPSRC) grant EP/S022139/1 - the Centre for Doctoral Training in Connected Electronic and Photonic Systems - and EPSRC Programme Grant TRANSNET EP/R035342/1. In addition, Polina Bayvel is supported through a Royal Society Research Professorship.

% Bibliography
\bibliography{references.bib}

%\onecolumn
%\centering
%\scriptsize

\newpage
\onecolumn
\appendix
\section{Appendix: Data from previous works and our benchmarks}
\label{appendix:A}
\nopagebreak
\setlength{\tabcolsep}{6pt}

\begin{longtable}[!htbp]{llll|lllll|l}

\caption*{Comparison of service blocking probabilities for KSP-FF and RL solutions across various topologies and traffic loads.}\label{tab:blocking_probabilities} \\
\hline
\multirow{3}{*}{\textbf{Publication}} & \multirow{3}{*}{\textbf{Topology}} & \multirow{3}{*}{\textbf{N$_{slots}$}} & \multirow{3}{*}{\textbf{\begin{tabular}[c]{@{}l@{}}Load \\ (Erlang)\end{tabular}}} & \multicolumn{5}{c|}{\textbf{Service blocking probability (\%)}} & \multirow{3}{*}{\textbf{\begin{tabular}[c]{@{}l@{}}Mean \\ improvement \\ over RL (\%)\end{tabular}}} \\
\cline{5-9}
 &  &  &  & \multicolumn{2}{c|}{Published} & \multicolumn{3}{c|}{Ours} &  \\
\cline{5-9}
 &  &  &  & RL & \multicolumn{1}{c|}{5-SP-FF} & 5-SP-FF & 5-SP-FF$_{hops}$ & 50-SP-FF$_{hops}$ &  \\
\hline
\endfirsthead

\multicolumn{10}{c}{continued from previous page} \\
\hline
\multirow{3}{*}{\textbf{Publication}} & \multirow{3}{*}{\textbf{Topology}} & \multirow{3}{*}{\textbf{N$_{slots}$}} & \multirow{3}{*}{\textbf{\begin{tabular}[c]{@{}l@{}}Load \\ (Erlang)\end{tabular}}} & \multicolumn{5}{c|}{\textbf{Service blocking probability (\%)}} & \multirow{3}{*}{\textbf{\begin{tabular}[c]{@{}l@{}}Mean \\ improvement \\ over RL (\%)\end{tabular}}} \\
\cline{5-9}
 &  &  &  & \multicolumn{2}{c|}{Published} & \multicolumn{3}{c|}{Ours} &  \\
\cline{5-9}
 &  &  &  & RL & \multicolumn{1}{c|}{5-SP-FF} & 5-SP-FF & 5-SP-FF$_{hops}$ & 50-SP-FF$_{hops}$ & \\
\hline
\endhead

\hline
\multicolumn{10}{r}{Continued on next page} \\
\endfoot

\hline
\endlastfoot

DeepRMSA             & NSFNET            & 100                  & 250                                                               & 4.00    & \multicolumn{1}{l|}{5.10}     & 5.00 ± 0.29 & 2.93 ± 0.22      & 2.33 ± 0.25                  & 42                                                                                    \\
DeepRMSA             & COST239           & 100                  & 600                                                               & 5.75 & \multicolumn{1}{l|}{6.75}    & 6.69 ± 0.35  & 3.80 ± 0.39      & 2.61 ± 0.36                  & 55                                                                                    \\
Reward-RMSA          & NSFNET            & 100                  & 168                                                               & 0.35  & \multicolumn{1}{l|}{1.00}       & 1.06 ± 0.11  & 0.10 ± 0.02      & 0.02 ± 0.01                  & 94                                                                                    \\
Reward-RMSA          & NSFNET            & 100                  & 182                                                               & 0.70  & \multicolumn{1}{l|}{1.40}       & 1.53 ± 0.15  & 0.25 ± 0.03      & 0.08 ± 0.03                  & 89                                                                                    \\
Reward-RMSA          & NSFNET            & 100                  & 196                                                               & 1.20  & \multicolumn{1}{l|}{2.10}       & 2.06 ± 0.12  & 0.48 ± 0.07      & 0.21 ± 0.04                  & 83                                                                                    \\
Reward-RMSA          & NSFNET            & 100                  & 210                                                               & 1.80  & \multicolumn{1}{l|}{2.70}       & 2.68 ± 0.20 & 0.95 ± 0.11      & 0.47 ± 0.05                  & 74                                                                                    \\                                                      
GCN-RMSA             & NSFNET            & 100                  & 154                                                               & 0.51 & \multicolumn{1}{l|}{0.67}    & 0.69 ± 0.08 & 0.03 ± 0.02      & 0.01 ± 0.00                  & 98                                                                                    \\
GCN-RMSA             & NSFNET            & 100                  & 168                                                               & 0.66 & \multicolumn{1}{l|}{1.00}       & 1.06 ± 0.11 & 0.10 ± 0.02      & 0.02 ± 0.01                  & 97                                                                                    \\
GCN-RMSA             & NSFNET            & 100                  & 182                                                               & 0.98 & \multicolumn{1}{l|}{1.42}    & 1.53 ± 0.15 & 0.25 ± 0.03      & 0.08 ± 0.03                  & 92                                                                                    \\
GCN-RMSA             & NSFNET            & 100                  & 196                                                               & 1.50  & \multicolumn{1}{l|}{2.10}     & 2.06 ± 0.12 & 0.48 ± 0.07      & 0.21 ± 0.04                  & 86                                                                                    \\
GCN-RMSA             & NSFNET            & 100                  & 210                                                               & 2.10  & \multicolumn{1}{l|}{2.76}    & 2.68 ± 0.20 & 0.95 ± 0.11      & 0.47 ± 0.05                  & 78                                                                                    \\
GCN-RMSA             & COST239           & 100                  & 368                                                               & 0.71 & \multicolumn{1}{l|}{0.78}    & 0.80 ± 0.08 & 0.07 ± 0.03      & 0.00 ± 0.00                  & 100                                                                                    \\
GCN-RMSA             & COST239           & 100                  & 391                                                               & 0.85 & \multicolumn{1}{l|}{1.30}     & 1.13 ± 0.12    & 0.17 ± 0.07      & 0.00 ± 0.00                  & 100                                                                                    \\
GCN-RMSA             & COST239           & 100                  & 414                                                               & 1.25 & \multicolumn{1}{l|}{1.70}     & 1.57 ± 0.13 & 0.28 ± 0.09      & 0.01 ± 0.01                  & 100                                                                                    \\
GCN-RMSA             & COST239           & 100                  & 437                                                               & 1.75 & \multicolumn{1}{l|}{2.30}     & 2.02 ± 0.13 & 0.46 ± 0.11      & 0.02 ± 0.02                  & 99                                                                                    \\
GCN-RMSA             & COST239           & 100                  & 460                                                               & 2.10  & \multicolumn{1}{l|}{2.70}     & 2.53 ± 0.21 & 0.73 ± 0.20      & 0.07 ± 0.03                  & 97                                                                                    \\
GCN-RMSA             & USNET             & 100                  & 320                                                               & 0.65 & \multicolumn{1}{l|}{0.75}    & 0.98 ± 0.10 & 0.25 ± 0.07      & 0.00 ± 0.01                  & 100                                                                                    \\
GCN-RMSA             & USNET             & 100                  & 340                                                               & 0.83 & \multicolumn{1}{l|}{1.20}     & 1.31 ± 0.16 & 0.42 ± 0.07      & 0.01 ± 0.02                  & 99                                                                                    \\
GCN-RMSA             & USNET             & 100                  & 360                                                               & 1.05 & \multicolumn{1}{l|}{1.50}     & 1.69 ± 0.12 & 0.73 ± 0.16      & 0.04 ± 0.04                  & 96                                                                                    \\
GCN-RMSA             & USNET             & 100                  & 380                                                               & 1.60  & \multicolumn{1}{l|}{2.15}    & 2.18 ± 0.19 & 1.05 ± 0.19      & 0.09 ± 0.09                  & 94                                                                                    \\
GCN-RMSA             & USNET             & 100                  & 400                                                               & 2.20  & \multicolumn{1}{l|}{2.70}     & 2.79 ± 0.24 & 1.53 ± 0.18      & 0.23 ± 0.09                  & 90                                                                                    \\
MaskRSA              & NSFNET            & 80                   & 80                                                                & 0.01 & \multicolumn{1}{l|}{0.26}    & 0.33 ± 0.06 & 0.00 ± 0.00      & 0.00 ± 0.00                  & 100                                                                                    \\
MaskRSA              & NSFNET            & 80                   & 90                                                                & 0.05 & \multicolumn{1}{l|}{0.60}     & 0.64 ± 0.07 & 0.01 ± 0.01      & 0.00 ± 0.00                  & 100                                                                                    \\
MaskRSA              & NSFNET            & 80                   & 100                                                               & 0.20  & \multicolumn{1}{l|}{0.90}     & 1.08 ± 0.10 & 0.08 ± 0.05      & 0.04 ± 0.02                  & 80                                                                                    \\
MaskRSA              & NSFNET            & 80                   & 110                                                               & 0.50  & \multicolumn{1}{l|}{1.80}     & 1.68 ± 0.11 & 0.25 ± 0.09      & 0.13 ± 0.03                  & 74                                                                                    \\
MaskRSA              & NSFNET            & 80                   & 120                                                               & 0.79 & \multicolumn{1}{l|}{2.47}    & 2.39 ± 0.14 & 0.60 ± 0.14      & 0.33 ± 0.12                  & 58                                                                                    \\
MaskRSA              & NSFNET            & 80                   & 130                                                               & 1.80  & \multicolumn{1}{l|}{3.00}       & 3.14 ± 0.22 & 1.35 ± 0.12      & 0.86 ± 0.14                  & 52                                                                                    \\
MaskRSA              & NSFNET            & 80                   & 140                                                               & 3.00    & \multicolumn{1}{l|}{4.00}       & 4.05 ± 0.21 & 2.26 ± 0.23      & 1.71 ± 0.27                  & 43                                                                                    \\
MaskRSA              & NSFNET            & 80                   & 150                                                               & 4.30  & \multicolumn{1}{l|}{5.00}       & 5.19 ± 0.28 & 3.43 ± 0.21      & 2.84 ± 0.26                  & 34                                                                                    \\
MaskRSA              & NSFNET            & 80                   & 160                                                               & 5.30  & \multicolumn{1}{l|}{6.00}       & 6.37 ± 0.26 & 4.65 ± 0.35      & 4.15 ± 0.27                  & 22                                                                                    \\
MaskRSA              & JPN48             & 80                   & 120                                                               & 0.05 & \multicolumn{1}{l|}{1.60}     & 1.92 ± 0.27 & 0.32 ± 0.08      & 0.00 ± 0.00                  & 100                                                                                    \\
MaskRSA              & JPN48             & 80                   & 130                                                               & 0.20  & \multicolumn{1}{l|}{2.50}     & 2.75 ± 0.31 & 0.64 ± 0.10      & 0.02 ± 0.01                  & 90                                                                                    \\
MaskRSA              & JPN48             & 80                   & 140                                                               & 0.55 & \multicolumn{1}{l|}{3.50}     & 3.69 ± 0.30 & 0.97 ± 0.14      & 0.04 ± 0.03                  & 93 \\
MaskRSA              & JPN48             & 80                   & 150                                                               & 0.90  & \multicolumn{1}{l|}{4.50}     & 4.55 ± 0.32 & 1.41 ± 0.15      & 0.09 ± 0.04                  & 90                                                                                    \\
MaskRSA              & JPN48             & 80                   & 160                                                               & 1.60  & \multicolumn{1}{l|}{5.00}       & 5.40 ± 0.36 & 2.00 ± 0.21      & 0.18 ± 0.04                  & 89                                                                                    \\
PtrNet-RSA           & NSFNET            & 40                   & 180                                                               & 0.01 & \multicolumn{1}{l|}{0.80}     & 0.93 ± 0.11 & 0.00 ± 0.00      & 0.00 ± 0.00                  & 100                                                                                    \\
PtrNet-RSA           & NSFNET            & 40                   & 190                                                               & 0.03 & \multicolumn{1}{l|}{1.25}    & 1.27 ± 0.10 & 0.01 ± 0.01      & 0.00 ± 0.00                  & 100                                                                                    \\
PtrNet-RSA           & NSFNET            & 40                   & 200                                                               & 0.08 & \multicolumn{1}{l|}{1.50}     & 1.68 ± 0.10 & 0.02 ± 0.01      & 0.00 ± 0.01                  & 100                                                                                    \\
PtrNet-RSA           & NSFNET            & 40                   & 210                                                               & 0.19 & \multicolumn{1}{l|}{1.90}     & 2.06 ± 0.11 & 0.09 ± 0.04      & 0.03 ± 0.02                  & 84                                                                                    \\
PtrNet-RSA           & NSFNET            & 40                   & 220                                                               & 0.23 & \multicolumn{1}{l|}{2.50}     & 2.58 ± 0.11 & 0.22 ± 0.09      & 0.09 ± 0.05                  & 61                                                                                    \\
PtrNet-RSA           & NSFNET            & 40                   & 230                                                               & 0.75 & \multicolumn{1}{l|}{2.90}     & 3.14 ± 0.14 & 0.48 ± 0.12      & 0.26 ± 0.10                  & 65                                                                                    \\
PtrNet-RSA           & NSFNET            & 40                   & 240                                                               & 1.30  & \multicolumn{1}{l|}{3.50}     & 4.01 ± 0.22 & 0.90 ± 0.18      & 0.54 ± 0.15                  & 58                                                                                    \\
PtrNet-RSA           & COST239           & 40                   & 340                                                               & 0.01 & \multicolumn{1}{l|}{2.50}     & 3.36 ± 0.16 & 0.02 ± 0.02      & 0.00 ± 0.00                  & 100                                                                                    \\
PtrNet-RSA           & COST239           & 40                   & 360                                                               & 0.04 & \multicolumn{1}{l|}{3.25}    & 4.02 ± 0.20  & 0.04 ± 0.02      & 0.00 ± 0.00                  & 100                                                                                    \\
PtrNet-RSA           & COST239           & 40                   & 380                                                               & 0.12 & \multicolumn{1}{l|}{3.80}     & 4.72 ± 0.29 & 0.10 ± 0.03      & 0.00 ± 0.00                  & 100                                                                                    \\
PtrNet-RSA           & COST239           & 40                   & 400                                                               & 0.24 & \multicolumn{1}{l|}{4.60}     & 5.43 ± 0.36 & 0.25 ± 0.09      & 0.00 ± 0.00                  & 100                                                                                    \\
PtrNet-RSA           & COST239           & 40                   & 420                                                               & 0.39 & \multicolumn{1}{l|}{5.50}     & 6.24 ± 0.25 & 0.46 ± 0.12      & 0.00 ± 0.00                  & 100                                                                                    \\
PtrNet-RSA           & USNET             & 40                   & 210                                                               & 0.01 & \multicolumn{1}{l|}{0.85}    & 1.02 ± 0.20 & 0.14 ± 0.05      & 0.00 ± 0.00                  & 100                                                                                    \\
PtrNet-RSA           & USNET             & 40                   & 220                                                               & 0.08 & \multicolumn{1}{l|}{1.40}     & 1.46 ± 0.25 & 0.29 ± 0.08      & 0.00 ± 0.01                  & 100                                                                                    \\
PtrNet-RSA           & USNET             & 40                   & 230                                                               & 0.22 & \multicolumn{1}{l|}{1.85}    & 1.99 ± 0.31 & 0.46 ± 0.10      & 0.02 ± 0.01                  & 91                                                                                    \\
PtrNet-RSA           & USNET             & 40                   & 240                                                               & 0.38 & \multicolumn{1}{l|}{2.20}     & 2.59 ± 0.38 & 0.77 ± 0.18      & 0.08 ± 0.06                  & 79                                                                                    \\
PtrNet-RSA           & USNET             & 40                   & 250                                                               & 0.68 & \multicolumn{1}{l|}{3.10}     & 3.25 ± 0.32 & 1.19 ± 0.28      & 0.19 ± 0.09                  & 72                                                                                    \\
PtrNet-RSA           & USNET             & 40                   & 260                                                               & 1.10  & \multicolumn{1}{l|}{3.60}     & 3.87 ± 0.49 & 1.81 ± 0.35      & 0.42 ± 0.15                  & 62                                                                                    \\
PtrNet-RSA           & USNET             & 40                   & 270                                                               & 1.80  & \multicolumn{1}{l|}{4.40}     & 4.67 ± 0.50 & 2.39 ± 0.41      & 0.72 ± 0.27                  & 60                                                                                    \\
PtrNet-RSA           & USNET             & 40                   & 280                                                               & 2.20  & \multicolumn{1}{l|}{4.90}     & 5.40 ± 0.46 & 3.16 ± 0.48      & 1.25 ± 0.34                  & 43                                                                                    \\
PtrNet-RSA           & NSFNET            & 80                   & 200                                                               & 0.01 & \multicolumn{1}{l|}{0.60}     & 0.61 ± 0.09 & 0.00 ± 0.00      & 0.00 ± 0.00                  & 100                                                                                    \\
PtrNet-RSA           & NSFNET            & 80                   & 210                                                               & 0.03 & \multicolumn{1}{l|}{0.80}     & 0.83 ± 0.10 & 0.00 ± 0.00      & 0.00 ± 0.00                  & 100                                                                                    \\
PtrNet-RSA           & NSFNET            & 80                   & 220                                                               & 0.11 & \multicolumn{1}{l|}{1.20}     & 1.07 ± 0.09 & 0.02 ± 0.02      & 0.01 ± 0.02                  & 91                                                                                    \\
PtrNet-RSA           & NSFNET            & 80                   & 230                                                               & 0.16 & \multicolumn{1}{l|}{1.40}     & 1.32 ± 0.13 & 0.05 ± 0.03      & 0.02 ± 0.02                  & 88                                                                                    \\
PtrNet-RSA           & NSFNET            & 80                   & 240                                                               & 0.50  & \multicolumn{1}{l|}{1.90}     & 1.63 ± 0.15 & 0.11 ± 0.06      & 0.05 ± 0.04                  & 90                                                                                    \\
PtrNet-RSA           & COST239           & 80                   & 420                                                               & 0.01 & \multicolumn{1}{l|}{2.50}     & 2.66 ± 0.11  & 0.08 ± 0.03      & 0.00 ± 0.00                  & 100                                                                                    \\
PtrNet-RSA           & COST239           & 80                   & 440                                                               & 0.16 & \multicolumn{1}{l|}{3.00}       & 3.05 ± 0.16 & 0.13 ± 0.06      & 0.00 ± 0.00                  & 100                                                                                    \\
PtrNet-RSA           & COST239           & 80                   & 460                                                               & 0.65 & \multicolumn{1}{l|}{3.50}     & 3.52 ± 0.19 & 0.25 ± 0.07      & 0.01 ± 0.01                  & 98                                                                                    \\
PtrNet-RSA           & USNET             & 80                   & 260                                                               & 0.03 & \multicolumn{1}{l|}{1.00}       & 1.21 ± 0.18 & 0.46 ± 0.13      & 0.00 ± 0.00                  & 100                                                                                    \\
PtrNet-RSA           & USNET             & 80                   & 270                                                               & 0.09 & \multicolumn{1}{l|}{1.20}     & 1.48 ± 0.24 & 0.64 ± 0.18      & 0.16 ± 0.09                  & -78                                                                                    \\
PtrNet-RSA           & USNET             & 80                   & 280                                                               & 0.15 & \multicolumn{1}{l|}{1.40}     & 1.78 ± 0.20 & 0.87 ± 0.21      & 0.38 ± 0.13                  & -153                                                                                    \\
PtrNet-RSA           & USNET             & 80                   & 290                                                               & 0.29 & \multicolumn{1}{l|}{1.70}     & 2.13 ± 0.24 & 1.12 ± 0.24      & 0.54 ± 0.15                  & -86                                                                                    \\
PtrNet-RSA           & USNET             & 80                   & 300                                                               & 0.52 & \multicolumn{1}{l|}{2.20}     & 2.46 ± 0.23 & 1.42 ± 0.29      & 0.61 ± 0.20                  & -33                                                                                    \\
PtrNet-RSA           & USNET             & 80                   & 310                                                               & 0.66 & \multicolumn{1}{l|}{2.40}     & 2.79 ± 0.26 & 1.75 ± 0.32      & 0.88 ± 0.23                  & -33                                                                                    \\
PtrNet-RSA           & USNET             & 80                   & 320                                                               & 1.00    & \multicolumn{1}{l|}{2.80}     & 3.18 ± 0.31 & 2.08 ± 0.35      & 1.15 ± 0.26                  & -15                                                                                    \\
PtrNet-RSA           & USNET             & 80                   & 330                                                               & 1.50  & \multicolumn{1}{l|}{3.10}     & 3.50 ± 0.28 & 2.42 ± 0.31      & 1.46 ± 0.30                  & 3                                                                                    
\end{longtable}

\end{document}